\documentclass[english,notitlepage,aps,prb,citeautoscript]{revtex4-1}
\usepackage[T1]{fontenc}
\usepackage[latin9]{inputenc}
\usepackage{amsmath}
\usepackage{amssymb}
\usepackage{graphicx}

\makeatletter

\providecommand{\tabularnewline}{\\}

\usepackage{babel}

\makeatother

\usepackage{babel}
\begin{document}
\title{Fast exchange with Gaussian basis set using robust pseudospectral
method}
\author{Sandeep Sharma}
\email{sanshar@gmail.com}

\affiliation{Department of Chemistry, University of Colorado Boulder, Boulder,
CO 80309}
\author{Alec F. White}
\email{white@qsimulate.com}

\affiliation{Quantum Simulation Technologies, Inc., Boston, 02135, United States}
\author{Gregory Beylkin}
\email{beylkin@colorado.edu}

\affiliation{Department of Applied Mathematics, University of Colorado, Boulder,
CO 80309, USA}
\begin{abstract}
In this article we present an algorithm to efficiently evaluate the
exchange matrix in periodic systems when Gaussian basis set with pseudopotentials
are used. The usual algorithm for evaluating exchange matrix scales
cubically with the system size because one has to perform $O(N^{2})$
fast Fourier transforms (FFT). Here we introduce an algorithm that
retains the cubic scaling but reduces the prefactor significantly
by eliminating the need to do FFTs during each exchange build. This
is accomplished by representing the products of Gaussian basis function
using a linear combination of an auxiliary basis the number of which
scales linearly with the size of the system. We store the potential
due to these auxiliary functions in memory which allows us to obtain
the exchange matrix without the need to do FFT, albeit at the cost
of additional memory requirement. Although the basic idea of using
auxiliary functions is not new, our algorithm is cheaper due to a
combination of three ingredients: (a) we use robust Pseudospectral
method that allows us to use a relatively small number of auxiliary
basis to obtain high accuracy (b) we use occ-RI exchange which eliminates
the need to construct the full exchange matrix and (c) we use the
(interpolative separable density fitting) ISDF algorithm to construct
these auxiliary basis that are used in the robust pseudospectral method.
The resulting algorithm is accurate and we note that the error in
the final energy decreases exponentially rapidly with the number of
auxiliary functions.  
\end{abstract}
\maketitle

\section{Introduction}

The inclusion of exact Hartree-Fock (HF) exchange within the framework
of Kohn-Sham (KS) density functional theory (DFT) is critical to the
success of DFT for molecular systems, and these ``hybrid'' functionals
are used in almost all modern DFT calculations on molecules. For periodic
solids, hybrid functionals can dramatically improve on standard semi-local
functionals for wide variety of properties \citep{Heyd2005,Kummel2008,Finazzi2008,Hai2011,Basera2019},
but the large computational cost associated with computing the HF
exchange is a limiting factor. The evaluation of the non-local HF
exchange can be computationally demanding in any context, and it is
particularly expensive for solids where calculations with hybrid density
functionals may be orders of magnitude more expensive than for their
purely semilocal counterparts.

Molecular calculations usually use local basis sets where the ratio
of basis functions to electrons, $N/n$, is small, often on the order
of 3-10. Though computation of the electron repulsion integrals is
naively $O(N^{4})$, the locality of the basis implies an asymptotically
linear number of significant basis function pairs and quadratic scaling
of the classical Coulomb and the HF exchange \citep{Almlof1982}.
The computation of the Coulomb energy and potential can be further
reduced to $O(N)$by multipole expansion \citep{White1994,Kudin1998},
while the exchange can be computed in linear time by leveraging the
sparsity of the density matrix for non-metallic systems \citep{Challacombe1997,Ochsenfeld1998,Ko2020,Goedecker1999}.
This asymptotically linear region is rarely reached in practice, and
recent work has focused on reducing the computational cost of practical
calculations by tensor factorization. The resolution of the identity
(RI) approximation \citep{Fruchtl1998,Weigend2002}, also called ``density
fitting,'' is the most widely used such method, and efficient approaches
for both Coulomb (RI-J) \citep{Sierka2003,Sodt2006} and exchange
(RI-K) \citep{Polly2004,Sodt2008,Manzer2015,Manzer2015a} have been
developed. Dunlap introduced a ``robust'' approximation to the 2-electron
integrals for which the error in integrals is quadratic in the fitting
error for the basis function products \citep{Dunlap2000,Dunlap2000a,Hollman2017}.
The Cholesky decomposition (CD) is another method that can be used
to obtain a decomposition of RI form without the need for optimized
auxiliary basis sets \citep{Beebe1977,Aquilante2007}. An alternative
approach, the pseudospectral (PS) method, is to provide a factorization
from a real-space perspective by introducing a basis of grid points
and performing one of the integrals analytically \citep{Friesner1985}.
The chain-of-spheres exchange (COSX) algorithm \citep{Neese2009}
and related semi-numerical exchange algorithms\citep{laqua18,Laqua20,Kong2017,kaup15}
are a particularly general application of the PS method to exchange.
The idea of factorizing the integral tensor was taken one step further
in the tensor hypercontraction (THC) method of Martinez and coworkers
where the 4-index 2-electron integral tensor is decomposed into a
product of five 2-index tensors \citep{Hohenstein2012,Parrish2012,Hohenstein2012a}.
The difficulty of efficiently performing the THC tensor decomposition
has largely limited its use to correlated methods, but a cubic scaling
factorization algorithm was first introduced by Lu and Ying \citep{Lu2015}
and later used to accelerate the computation of exact exchange in
periodic \citep{Hu2017,Dong2018} and molecular \citep{Lee2020} calculations
under the name ``interpolative separable density fitting'' (ISDF).

In traditional plane-wave DFT calculations, the action of the Coulomb
operator on the occupied orbitals can be evaluated by solving $n$
Poisson equations. This scales quadratically in total, ($O(nN_{g}\ln N_{g})$
for $n$ electrons and $N_{g}$ grid points). In contrast, the action
of the exchange operator scales cubically ($O(n^{2}N_{g}\ln N_{g})$
for $n$ electrons and $N_{g}$ grid points), a fact that has motivated
the development of numerous numerical methods that seek to lessen
this cost. Most notable are linear scaling methods \citep{Goedecker1999,Bowler2012}
for which the sparsity in the exchange operator relies on the system
being an insulator \citep{Wu2009}. Stochastic density functional
theory (sDFT) \citep{Baer2013}, including the extension to hybrid
functionals \citep{Neuhauser2016}, can achieve linear scaling without
any locality arguments, but controlling the stochastic error results
in a very large prefactor. As with molecular systems, the linear regime
is usually out-of-reach, and methods that do not improve the scaling,
like the adaptively compressed exchange (ACE) \citep{Lin2016}, can
greatly increase the efficiency in practice. When large supercells
are necessary, a screening and/or truncating the coulomb operator
in the exchange term can somewhat lessen the cost \citep{Todorova2006}.
The auxiliary density matrix method \citep{Guidon2010} (ADMM) can
significantly reduce the cost by approximating the density matrix
\citep{Guidon2010}.

In this work, we present an efficient algorithm for evaluating the
exchange matrix in periodic Gaussian type orbital (GTO) calculations.
We begin by providing background information that will help the reader
understand the reason for high cost of exchange matrix evaluation,
namely that one has to perform $O(N^{2})$ FFTs. In Section~\ref{sec:Speeding-up-exchange}
we describe the algorithm which combines several different ideas that
eliminate the need for performing FFTs during exchange build and replace
them with matrix multiplications. Although the algorithm is still
cubic scaling the prefactor is significantly reduced. In Section~\ref{sec:Computational-details}
we describe the computational details including an efficient parallel
implementation. Finally, in Section~\ref{sec:Results} we show that
this algorithm enables computation of the exchange matrix with a cost
comparable to the computation of the Coulomb matrix. This allows hybrid
DFT to be used for systems where semi-local DFT is feasible.

\section{Background\label{sec:Background}}

\begin{table}
\begin{centering}
{\small{}}%
\begin{tabular}{ccl}
\hline 
{\small{}Notation } &  & {\small{}Meaning}\tabularnewline
\hline 
{\small{}$n$ } &  & {\small{}number of electrons}\tabularnewline
{\small{}$N$ } &  & {\small{}number of atom centered Gaussian basis functions}\tabularnewline
{\small{}$N_{g}$ } &  & {\small{}number of grid points or plane wave basis}\tabularnewline
{\small{}$N_{\chi}$ } &  & {\small{}number of grid points used in pseudo-spectral method}\tabularnewline
{\small{}$k$ } &  & {\small{}number of processors (including MPI and OMP)}\tabularnewline
{\small{}$k_{t}$} &  & {\small{}number of threads per node}\tabularnewline
{\small{}$\phi_{i}$ or with subscripts $i,j,k,l,\cdots$ } &  & {\small{}occupied molecular orbitals}\tabularnewline
{\small{}$\phi_{\mu}$ or with subscripts $\mu,\nu,\lambda,\sigma\cdots$ } &  & {\small{}atom centered gaussian basis}\tabularnewline
{\small{}$\chi$ } &  & {\small{}pseudo-spectral fitting functions}\tabularnewline
{\small{}$\mathbf{R}$ } &  & {\small{}location of the grid point (or the center of the periodized
Sinc basis)}\tabularnewline
{\small{}$\mathbf{G}$ } &  & {\small{}wave-number of the plane wave}\tabularnewline
\hline 
\end{tabular}{\small\par}
\par\end{centering}
\caption{Notation that will be used in the rest of the paper\label{tab:Notation-that-will}}
\end{table}

In this work we will solve the Hartree-Fock equations using the self-consistent
field (SCF) method, where by, for a given set of molecular orbitals
one constructs the Coulomb ($\hat{J}$) and the Exchange ($\hat{K}$)
operators respectively. The Coulomb and exchange operators are given
by the expressions 
\begin{align*}
\hat{J}(\mathbf{r}) & =\int_{\mathbb{R}^{3}}\sum_{i}\frac{|\phi_{i}(\mathbf{r}')|^{2}}{|\mathbf{r}-\mathbf{r}'|}d\mathbf{r}',\\
\hat{K}(\mathbf{r},\mathbf{r}') & =\sum_{i}\frac{\phi_{i}(\mathbf{r})\phi_{i}(\mathbf{r}')}{|\mathbf{r}-\mathbf{r}'|},
\end{align*}
where $\phi_{i}(\mathbf{r})$ are the occupied molecular orbitals.
From the equation one can notice that while the Coulomb operator is
diagonal, the exchange operator has a rank $n$. To make progress
one typically introduces a basis set and obtains the Coulomb and exchange
matrices $J$ and $K$ respectively. Although many possible basis
functions can be used including Slater type orbitals \citep{Cohen2002,Chong2004},
wavelets \citep{Goedecker2009,yanai2015,Genovese2008,White2019,mrchem},
numerical basis functions \citep{Blum2009,octopus2015,C5CP00110B,Enkovaara_2010}
etc. the two most commonly used ones are the atom-centered Gaussian
basis functions \citep{Ye2021,Irmler2018,Kuhne2020,Paier2009,Causa1988,Causa1988a,Kudin2004,Kudin2000,Sun2020,Lee2021}
and the plane-wave basis \citep{Kresse1996,Kresse1996a,Giannozzi_2017,GONZE2002478,holzwarth2005}.
Although, in this work we will use the Gaussian basis functions to
represent molecular orbitals, we will also make use of the plane wave
basis and their dual basis (the periodized Sinc basis) to simplify
certain calculations \citep{pulay,pulay2,Sun2017,Vandevondele2005}.
We first review properties of these functions.

\subsection{Basis functions}

We begin by introducing periodic basis functions in the unit cell
defined by vectors $\mathbf{a}_{1},\mathbf{a}_{2},\mathbf{a}_{3}$,
such that the volume $\Omega=\mathbf{a}_{1}\cdot(\mathbf{a}_{2}\times\mathbf{a}_{3})$.
The reciprocal vectors are denoted by $\mathbf{A}_{1},\mathbf{A}_{2},\mathbf{A}_{3}$
so that $\mathbf{a}_{i}\cdot\mathbf{A}_{j}=2\pi\delta_{ij}$.

\subsubsection{Plane wave basis}

The plane wave basis functions $\left(\xi_{\mathbf{G}}\right)$ are
parameterized by the wave-vector $\mathbf{G}$ and are given by 
\[
\xi_{\mathbf{G}}\left(\mathbf{r}\right)=\frac{1}{\sqrt{\Omega}}\exp\left(-i\mathbf{G}\cdot\mathbf{r}\right).
\]
It is easy to check that these functions form an orthogonal basis
i.e. 
\[
\int_{\Omega}\xi_{\mathbf{G'}}(\mathbf{r})^{*}\xi_{\mathbf{G}}(\mathbf{r})d\mathbf{r}=\delta_{\mathbf{G',\mathbf{G}}}
\]
as long as $\mathbf{G}=n_{1}\mathbf{A}_{1}+n_{2}\mathbf{A}_{2}+n_{3}\mathbf{A}_{3}$,
where $n_{i}$ are integers. If in a calculation one retains $n_{i}=-\frac{N_{i}}{2},\cdots,\frac{N_{i}+1}{2}$,
where $N_{i}$ is an integer, $i=1,2,3$, then the total number of
plane waves is equal to $N_{g}=N_{1}N_{2}N_{3}$.

\subsubsection{Sinc basis}

The dual basis consists of the periodized Sinc functions $\left(\xi_{\mathbf{R}}\right)$
(referred to as pSinc functions from now on) which are associated
with grid points $\mathbf{R}$. They are obtained by the unitary transformation
of the plane wave basis functions 
\begin{align}
\xi_{\mathbf{R}}\left(\mathbf{r}\right) & =\frac{1}{\sqrt{N_{g}}}\sum_{\mathbf{G}}\exp\left(i\mathbf{G}\cdot\mathbf{R}\right)\xi_{\mathbf{G}}\left(\mathbf{r}\right)\label{eq:PS}\\
\xi_{\mathbf{G}}\left(\mathbf{r}\right) & =\frac{1}{\sqrt{N_{g}}}\sum_{\mathbf{G}}\exp\left(-i\mathbf{G}\cdot\mathbf{R}\right)\xi_{\mathbf{R}}\left(\mathbf{r}\right)
\end{align}
Again it is easy to see that $\xi_{\mathbf{R}}$ are orthogonal if
$\mathbf{R}=\frac{n_{1}\mathbf{a}_{1}}{N_{1}}+\frac{n_{2}\mathbf{a}_{2}}{N_{2}}+\frac{n_{3}\mathbf{a}_{3}}{N_{3}}$
are a uniform set of grid points in the unit cell. Since the plane
wave and the pSinc bases are related via a unitary transformation,
they span the same space.

\subsubsection{Atom-centered Gaussian basis}

We also have the computational basis, the periodized atom centered
Gaussian functions $\phi_{P\mu}(\mathbf{r})$ given by 
\[
\phi_{P\mu}\left(\mathbf{r}\right)=\sum_{\mathbf{T}}\phi_{\mu}\left(\mathbf{r}+\mathbf{T}\right),
\]
where $\phi\left(\mathbf{r}\right)$ is the atom centered Gaussian
and $\mathbf{T}=n_{1}\mathbf{a}_{1}+n_{2}\mathbf{a}_{2}+n_{3}\mathbf{a}_{3}$
are the lattice vectors in the real space. This function is also known
as a Jacobi theta function. Note that $\phi_{P\mu}\left(\mathbf{r}\right)$
is neither orthogonal or normalized even if the functions $\phi\left(\mathbf{r}\right)$
themselves are. The periodized Gaussians can be represented as a linear
combination of the plane waves , 
\begin{equation}
\phi_{P,\mu}\left(\mathbf{r}\right)=\sum_{\mathbf{G}}\hat{\phi}_{\mu}\left(\mathbf{G}\right)\xi_{\mathbf{G}}\left(\mathbf{r}\right)\label{eq:gaussPW}
\end{equation}
where $\hat{\phi}_{\mu}\left(\mathbf{G}\right)$ is the Fourier transform
of the function $\phi_{\mu}\left(\mathbf{r}\right)$, 
\[
\hat{\phi}_{\mu}\left(\mathbf{G}\right)=\int_{\mathbb{R}^{3}}\exp\left(i\mathbf{G}\cdot\mathbf{r}\right)\phi_{\mu}\left(\mathbf{r}\right)d\mathbf{r}
\]
The Fourier transform of a Gaussian basis function is also a Gaussian,
thus if 
\[
\phi_{\mu}(\mathbf{r})=\exp(-\mu|\mathbf{r}-\mathbf{A}_{\mu}|^{2})
\]
then 
\begin{equation}
\hat{\phi}_{\mu}(\mathbf{G})=\left(\frac{\pi}{\mu}\right)^{3/2}\frac{\exp(-|\mathbf{G}|^{2}/4/\mu)}{\sqrt{\Omega}}\exp(i\mathbf{G}\cdot\mathbf{A}_{\mu})\label{eq:gaussG}
\end{equation}
If the factor $\mu$ of the exponent of the Gaussian is small then
a relatively few plane waves $N_{g}$ are needed to represent it (with
a small but finite error). In this work we will use pseudo-potentials
so that the Gaussian basis functions will be reasonably flat and a
manageable number of plane waves is sufficient to represent them.

It is also possible to represent the periodized Gaussians as a linear
combination of the pSinc basis functions, 
\begin{align*}
\phi_{P\mu}\left(\mathbf{r}\right) & =\sum_{\mathbf{R}}\tilde{\phi}_{\mu}\left(\mathbf{R}\right)\xi_{\mathbf{R}}\left(\mathbf{r}\right)\\
\tilde{\phi}_{\mu}\left(\mathbf{R}\right) & =\frac{1}{\sqrt{N_{g}}}\sum_{\mathbf{G}}\exp\left(-i\mathbf{G}\cdot\mathbf{R}\right)\hat{\phi}_{\mu}\left(\mathbf{G}\right)\\
\hat{\phi}_{\mu}\left(\mathbf{G}\right) & =\frac{1}{\sqrt{N_{g}}}\sum_{\mathbf{R}}\exp\left(i\mathbf{G}\cdot\mathbf{R}\right)\tilde{\phi}_{\mu}\left(\mathbf{R}\right)
\end{align*}
These equations follow from (\ref{eq:gaussPW}) and (\ref{eq:PS})
respectively. In what follows it will be convenient to switch from
the pSinc basis to the plane-wave basis using the Fast Fourier transform
(FFT), i.e. the coefficients $\tilde{\phi}_{\mu}(\mathbf{R})$ can
be calculated from those of $\hat{\phi}_{\mu}(\mathbf{G})$ (and vice
versa) rapidly using FFT at a cost of $O(N_{g}\ln N_{g})$.

So far we have been careful to distinguish between $\phi_{\mu}$ and
$\phi_{P\mu}$, however, we will not do so for the rest of the paper
where we only use the symbol $\phi_{\mu}$ and it should be understood
that it refers to the periodized Gaussian function.

\subsection{Integral evaluation and the diagonal approximation}

In what follows we evaluate the potential due to a charge density
where the charge density is given as a product of two orbitals. If
we represent both orbitals $\phi_{\mu}\left(\mathbf{r}\right)$ and
$\phi_{\nu}\left(\mathbf{r}\right)$ using plane wave basis, we then
obtain the charge density $\rho_{\mu\nu}(\mathbf{r})$ as 
\begin{align*}
\rho_{\mu\nu}(\mathbf{r}) & =\phi_{\mu}(\mathbf{r})\phi_{\nu}(\mathbf{r})\\
 & =\frac{1}{\sqrt{\Omega}}\sum_{\mathbf{G}}^{N_{g}}\sum_{\mathbf{G}'}^{N_{g}}\hat{\phi}_{\mu}(\mathbf{G})\hat{\phi}_{\nu}(\mathbf{G}')\xi_{\mathbf{G}+\mathbf{G}'}(\mathbf{r})\\
 & =\frac{1}{\sqrt{\Omega}}\sum_{\mathbf{G}}^{2N_{g}}\sum_{\mathbf{G}'}^{N_{g}}\phi_{\mu}(\mathbf{G}-\mathbf{G}')\hat{\phi}_{\nu}(\mathbf{G}')\xi_{\mathbf{G}}(\mathbf{r})\\
 & \approx\frac{1}{\sqrt{\Omega}}\sum_{\mathbf{G}}^{N_{g}}\sum_{\mathbf{G}'}^{N_{g}}\phi_{\mu}(\mathbf{G}-\mathbf{G}')\hat{\phi}_{\nu}(\mathbf{G}')\xi_{\mathbf{G}}(\mathbf{r}),
\end{align*}
where in going from the second to the third expression we have changed
the variables and in the last expression we made an approximation
by truncating the summation over $\mathbf{G}$. By choosing a sufficiently
large cutoff $N_{g}$, the error due to the approximation can be made
negligibly small since we expect the density to be sufficiently smooth
so that the contributions to $\rho_{\mu\nu}(\mathbf{G})$ coming from
plane waves above the cutoff are small. With this approximation the
final equation starts to look like a convolution and one can show
that 
\begin{align}
\rho_{\mu\nu}(\mathbf{r}) & \approx\frac{1}{\sqrt{\Omega}}\sum_{\mathbf{G}}^{N_{g}}\sum_{\mathbf{G}'}^{N_{g}}\phi_{\mu}(\mathbf{G}-\mathbf{G}')\hat{\phi}_{\nu}(\mathbf{G}')\xi_{\mathbf{G}}(\mathbf{r})\nonumber \\
 & \approx\sqrt{\frac{N_{g}}{\Omega}}\sum_{\mathbf{\mathbf{R}}}^{N_{g}}\tilde{\phi}_{\mu}(\mathbf{R})\tilde{\phi}_{\nu}(\mathbf{R})\xi_{\mathbf{R}}(\mathbf{r})\label{eq:convol}
\end{align}
The last expression has been called the diagonal approximation and
has been extensively used in the work of Steve White in the context
of Gausslets \citep{White2019,Qiu2021,White2017} and before that
with Sinc functions \citep{Jones2016} .

Using the diagonal approximation described above we can calculate
the two-electron integrals of four Gaussian basis functions as follows
\begin{align}
(\mu\nu|\lambda\sigma) & =\int\int\phi_{\mu}(\mathbf{r})\phi_{\nu}(\mathbf{r})\frac{1}{|\mathbf{r}-\mathbf{r}'|}\phi_{\lambda}(\mathbf{r}')\phi_{\sigma}(\mathbf{r}')d\mathbf{r}d\mathbf{r'}\nonumber \\
 & =\frac{N_{g}}{\Omega}\sum_{\mathbf{R}}\sum_{\mathbf{R}'}\tilde{\phi}_{\mu}(\mathbf{R})\tilde{\phi}_{\nu}(\mathbf{R})\tilde{\phi}_{\lambda}(\mathbf{R}')\tilde{\phi}_{\sigma}(\mathbf{R}')\int\int\xi_{\mathbf{R}}(\mathbf{r})\frac{1}{|\mathbf{r}-\mathbf{r}'|}\xi_{\mathbf{R}'}(\mathbf{r}')d\mathbf{r}d\mathbf{r'}\nonumber \\
 & =\frac{N_{g}}{\Omega}\sum_{\mathbf{R}}\sum_{\mathbf{R}'}\tilde{\phi}_{\mu}(\mathbf{R})\tilde{\phi}_{\nu}(\mathbf{R})v(\mathbf{R}-\mathbf{R}')\tilde{\phi}_{\lambda}(\mathbf{R}')\tilde{\phi}_{\sigma}(\mathbf{R}')\label{eq:2eint}\\
v(\mathbf{R}-\mathbf{R}') & =\int\int\xi_{\mathbf{R}}(\mathbf{r})\frac{1}{|\mathbf{r}-\mathbf{r}'|}\xi_{\mathbf{R}'}(\mathbf{r}')d\mathbf{r}d\mathbf{r'}\label{eq:cou}\\
 & =\frac{1}{N_{g}}\sum_{\mathbf{G}}\exp(-i\mathbf{G}\cdot\mathbf{R})\frac{4\pi}{G^{2}}\exp(i\mathbf{G}\cdot\mathbf{R}')\label{eq:cou-1}
\end{align}

In going from the first to the second expression in \eqref{eq:2eint}
we have used (\ref{eq:convol}) and in going from the second to the
third expression we have used (\ref{eq:PS}) and made use of the fact
that the Coulomb operator is diagonal and is equal to $\frac{4\pi}{|\mathbf{G}|^{2}}$
in plane wave basis. It is useful to note that Equation (\ref{eq:2eint})
takes the form of tensor hypercontraction (THC) and just like in THC
one can make use of the integrals in this form to calculate exchange
with a cubic scaling cost. As we will show next, the Coulomb matrix
can be evaluated using a linear scaling algorithm because the matrix
$v(\mathbf{R}-\mathbf{R}')$ takes a special form shown in (\ref{eq:cou}).

\subsection{Coulomb matrix}

In the self-consistent field calculation we construct the coulomb
matrix $J_{\mu\nu}=\langle\mu|\hat{J}|\nu\rangle$ for a set of periodized
Gaussian functions $\phi_{\mu}$ and $\phi_{\nu}$. Further, we represent
the molecular orbitals as a linear combination of the basis functions
$\phi_{i}(\mathbf{r})=\sum_{\lambda}C_{\lambda i}\phi_{P\lambda}(\mathbf{r})$,
where $C_{\lambda i}$ is the matrix of molecular coefficients and
we also define the density matrix $D_{\lambda\sigma}=\sum_{i}C_{\lambda i}C_{\sigma i}$.
Using these, the Coulomb matrix can be written as 
\begin{align*}
J_{\mu\nu} & =\int d\mathbf{r}'\phi_{\mu}(\mathbf{r}')\phi_{\nu}(\mathbf{r}')\left(\int d\mathbf{r}\frac{1}{|\mathbf{r}-\mathbf{r}'|}\sum_{\lambda\sigma}D_{\lambda\sigma}\phi_{\lambda}(\mathbf{r})\phi_{\sigma}(\mathbf{r})\right)\\
 & =\frac{1}{\Omega}\sum_{\mathbf{R}}\left(\tilde{\phi}_{\mu}(\mathbf{R})\tilde{\phi}_{\nu}(\mathbf{R})\right)\left(\sum_{\mathbf{G}}\exp(-i\mathbf{G}\cdot\mathbf{R})\frac{4\pi}{G^{2}}\left(\sum_{\mathbf{R}'}\exp(i\mathbf{G}\cdot\mathbf{R}')\left(\sum_{\lambda\sigma}D_{\lambda\sigma}\tilde{\phi}_{\lambda}(\mathbf{R}')\tilde{\phi}_{\sigma}(\mathbf{R}')\right)\right)\right)
\end{align*}
where we have made use of (\ref{eq:2eint}) and (\ref{eq:cou}). Also
the order in which the brackets are placed specify the order in which
the tensor contractions are performed. Specifically, the steps involved
are 
\begin{enumerate}
\item We first evaluate the density on a grid $\rho(\mathbf{R}')=\sum_{\lambda\sigma}D_{\lambda\sigma}\tilde{\phi}_{\lambda}(\mathbf{R}')\tilde{\phi}_{\sigma}(\mathbf{R}')$.
This can be performed with a linear cost, since for any finite threshold,
the atom centered Gaussian basis functions have a compact support,
which implies that for a given $\lambda$ only an $\mathcal{O}\left(1\right)$
number of $\sigma$ have non-zero overlap. 
\item Next the density in Fourier space is evaluated $\rho(\mathbf{G})=\sum_{\mathbf{R}'}\exp(i\mathbf{G}\cdot\mathbf{R}')\rho(\mathbf{R}')$
which is done efficiently with $N_{g}\ln N_{g}$ cost using the FFT. 
\item In the next step the potential due to this density is evaluated in
real space as $V(\mathbf{R}')=\sum_{\mathbf{G}}\exp(-i\mathbf{G}\cdot\mathbf{R})\frac{4\pi}{G^{2}}\rho(\mathbf{G})$
which can also be evaluated efficiently using FFT with an $N_{g}\ln N_{g}$
cost. 
\item Finally, the Coulomb matrix is evaluated as $J_{\mu\nu}=\sum_{\mathbf{R}}\tilde{\phi}_{\mu}(\mathbf{R})\tilde{\phi}_{\nu}(\mathbf{R})V(\mathbf{R}')$,
which again is calculated with a linear scaling due to the compact
support of the Gaussian basis set. 
\end{enumerate}
Thus, the construction of the Coulomb matrix only requires a two FFT
and, the step 4 is the dominant cost of the calculation.

\subsection{Exchange matrix}

The exchange matrix $\langle\mu|\hat{K}|\nu\rangle$ can be constructed
in a number of ways using a quartic scaling algorithm e.g. by using
density fitting. However a cubic scaling algorithm is as follows,
\begin{align}
K_{\mu\nu} & =\sum_{i}\int\phi_{\mu}(\mathbf{r}')\phi_{i}(\mathbf{r})\left(\int\frac{\phi_{i}(\mathbf{r}')\phi_{\nu}(\mathbf{r}')}{|\mathbf{r}-\mathbf{r}'|}d\mathbf{r}'\right)d\mathbf{r}\nonumber \\
 & =\frac{1}{\Omega}\sum_{\mathbf{R}}\left(\tilde{\phi}_{\mu}(\mathbf{R})\tilde{\phi}_{i}(\mathbf{R})\right)\left(\sum_{\mathbf{G}}\exp(-i\mathbf{G}\cdot\mathbf{R})\frac{4\pi}{G^{2}}\left(\sum_{\mathbf{R}'}\exp(i\mathbf{G}\cdot\mathbf{R}')\tilde{\phi}_{i}(\mathbf{R}')\tilde{\phi}_{\nu}(\mathbf{R}')\right)\right),\label{eq:exchange}
\end{align}
where the order of tensor contraction is represented by the brackets.
The steps involved in constructing the exchange matrix are 
\begin{enumerate}
\item We first evaluate the product of the molecular orbital $\tilde{\phi}_{i}(\mathbf{R}')$
and atomic orbital $\tilde{\phi}_{\nu}(\mathbf{R}')$ on the grid
to obtain $\tilde{\rho}_{i\nu}(\mathbf{R}')=\tilde{\phi}_{i}(\mathbf{R}')\tilde{\phi}_{\nu}(\mathbf{R}')$,
which is an $\mathcal{O}\left(1\right)$ operation for a given $i,\nu$.
The value of this product density is then evaluated in the Fourier
space $\tilde{\rho}_{i\nu}(\mathbf{G})$ using FFT. 
\item Next the Coulomb kernel $\frac{4\pi}{G^{2}}$ multiplies $\tilde{\rho}_{i\nu}(\mathbf{G})$
and the inverse FFT is used to evaluate the potential $\tilde{V}_{i\nu}(\mathbf{R}')$
due to $\tilde{\rho}_{i\nu}(\mathbf{R}')$. 
\item Finally, the potential is contracted with the product $\tilde{\phi}_{\mu}(\mathbf{R})\tilde{\phi}_{i}(\mathbf{R})$
to obtain the value of $K_{\mu\nu}$. 
\end{enumerate}
Thus one has to perform $Nn$ FFTs and inverse FFTs, one pair each
for a given $i$ and $\nu$. The cost of the FFT is $N_{g}\ln N_{g}$,
which makes the total cost of the exchange evaluations equal to $NnN_{g}\ln N_{g}$,
where, as explained in Table~\ref{tab:Notation-that-will}, $N,n$
and $N_{g}$ are respectively the number of basis functions, number
of electrons (or occupied orbitals) and number of grid points (or
plane wave basis) respectively.

\subsection{Other steps}

For performing the HF calculation, one also needs the Kinetic matrix
$T_{\mu\nu}$, the nuclear matrix $N_{\mu\nu}$ and the overlap matrix
$S_{\mu\nu}$. These matrices are evaluated using standard techniques
\citep{sharmaBeylkina} and we do not discuss them further. Finally,
the Fock matrix $F_{\mu\nu}=T_{\mu\nu}+N_{\mu\nu}+J_{\mu\nu}-K_{\mu\nu}$
is constructed and diagonalized to obtain the molecular coefficients
$C_{\lambda i}$ using which the molecular orbitals $\phi_{i}(\mathbf{r})$
are formed. These molecular orbitals are used to construct the Fock
matrix in a self-consistent field cycle. The diagonalization of the
Fock matrix to obtain the molecular orbitals scales cubically, however
the prefactor of this step is small so that its cost does not exceed
that of Coulomb matrix construction (which scales linearly) for systems
with less than 5000 basis functions (for example, see Table~\ref{tab:All-calculations-are}).

\section{Speeding up exchange evaluation\label{sec:Speeding-up-exchange}}

As we mentioned in the previous section, the exchange evaluation scales
cubically with the size of the system. In particular, we have to perform
$Nn$ FFTs. In practice, the cubic scaling in itself is not a severe
limitation (at least for problems in which $\leq$5000 basis functions
are utilized) since other steps such as Fock matrix diagonalization
scale cubically as well. However, the high prefactor associated with
performing $Nn$ FFTs results in a large overall cost. In principle,
one can make evaluation of the exchange matrix to scale linearly by
using the fact that $\hat{K}(\mathbf{r},\mathbf{r}')$ decays exponentially
with $|\mathbf{r}-\mathbf{r}'|$ for systems with a band-gap or metallic
systems at finite temperatures. However, even for insulating systems
with reasonably large band gap the exponential decay is extremely
slow and one does not reach the linear scaling regime until very large
system sizes. In this work we take the point of view that by reducing
the prefactor in the cubic scaling algorithm, we can obtain a method
for exchange matrix formation that is only slightly more expensive
than the formation of the Coulomb matrix. We briefly describe how
this is accomplished before going into details in subsequent subsections.

Let us begin by observing that the number of pairs of basis function
($\phi_{\mu}\phi_{\nu}$) scale quadratically with the size of the
system. However, (\ref{eq:convol}) shows that by using the diagonal
approximation (which can be made arbitrarily accurate) one can ensure
that the products can be represented by using a linear combination
of pSinc basis the number of which scales linearly with the size of
the system. This fact reduces the scaling of the exchange build to
cubic (from the usual quartic), however with a large prefactor. We
will retain the cubic scaling but reduce the prefactor by fitting
the products of gaussians using a linear combination of relatively
small number of auxiliary functions. We are able to get away with
using a small number of auxiliary functions because we make use of
the so called robust pseudospectral method, which ensures that the
error in the exchange matrix is quadratic of the error in the fitting.
Below we first show how the auxiliary functions (represented by $\chi_{g}$)
are obtained and then how they are used in robust pseudospectral method.
Finally, we also introduce occ-RI exchange algorithm that further
allows us to reduce the cost.

\subsection{Interpolative separable density fitting}

This algorithm was introduced by Lu and Ying \citep{Lu2015} for writing
down the two electron integrals in the THC format and later it was
adapted for speeding up exchange matrix evaluation in hybrid DFT with
plane wave basis \citep{Hu2017,Dong2018}. The basis idea is to view
the product density 
\[
\tilde{\rho}_{\mu\nu,\mathbf{R}}=\tilde{\phi}_{\mu}(\mathbf{R})\tilde{\phi}_{\nu}(\mathbf{R})
\]
as a matrix with $N^{2}$ rows corresponding to indices $\mu\nu$
and $N_{g}$ columns corresponding to the grid points. One then tries
to obtain a subset of columns such that all other columns can be written
as a linear combination of this subset. Specifically, we will attempt
to write 
\begin{equation}
\tilde{\rho}_{\mu\nu,\mathbf{R}}=\tilde{\phi}_{\mu}(\mathbf{R})\tilde{\phi}_{\nu}(\mathbf{R})\approx\sum_{\mathbf{R}_{g}}\tilde{\phi}_{\mu}(\mathbf{R}_{g})\tilde{\phi}_{\nu}(\mathbf{R}_{g})\chi_{g}(\mathbf{R})+O(\epsilon),\label{eq:isdf}
\end{equation}
where $\mathbf{R}_{g}$ are a subset of all grid points that we call
interpolation points, $\chi_{g}(\mathbf{R})$ is a function defined
on all grid points and $\epsilon$ is the error incurred due to the
approximation. The interpolation grid points $\mathbf{R}_{g}$ are
obtained by performing QR decomposition with column pivoting (QRCP)\citep{engler1997}
on the matrix $\tilde{\rho}_{\mu\nu,\mathbf{R}}$, which allows one
to write it as 
\[
\tilde{\rho}P=QR
\]
where $P$ is a permutation matrix that ensures that the diagonal
entries of the matrix $R$ are in the descending order,$R_{11}\geq R_{22}\geq R_{33}\cdots$.
We then choose a user-defined $N_{\chi}$ set of interpolation points
specified by the leading pivot points in $P$. These provide an optimal
set of interpolation points. The cost of performing this step directly
scales as $N^{2}N_{g}^{2}$ (assuming $N^{2}>N_{g}$) and is thus
prohibitively expensive. It can be made to scale cubically by using
a randomized algorithm and we provide more details in the Section
\ref{subsec:ISDF}.

Once the interpolation points are selected we can obtain the function
$\chi_{g}(\mathbf{R})$ using a least square minimization 
\[
\min_{\chi_{\mathbf{R}_{g},\mathbf{R}}}\left|\tilde{\rho}_{\mu\nu,\mathbf{R}}-\tilde{\rho}_{\mu\nu,\mathbf{R}_{g}}\chi_{\mathbf{R}_{g},\mathbf{R}}\right|_{F},
\]
where we have written $\chi_{g}(\mathbf{R})$ as a matrix $\chi_{\mathbf{R}_{g},\mathbf{R}}$,
used the Einstein notation of summing over repeated indices (which
we will continue to do so for the rest of this subsection) and subscript
$F$ indices the Frobenius norm. The least square problem is solved
by transforming it into a system of linear equations 
\begin{align}
\tilde{\rho}_{\mu\nu,\mathbf{R}'_{g}}\tilde{\rho}_{\mu\nu,\mathbf{R}_{g}}\chi_{\mathbf{R}_{g},\mathbf{R}} & =\tilde{\rho}_{\mu\nu,\mathbf{R}'_{g}}\tilde{\rho}_{\mu\nu,\mathbf{R}}\nonumber \\
X_{\mathbf{R}'_{g},\mathbf{R}_{g}}\chi_{\mathbf{R}_{g},\mathbf{R}} & =X_{\mathbf{R}'_{g},\mathbf{R}},\label{eq:lsqfit}
\end{align}
where $X=\tilde{\rho}^{T}\tilde{\rho}$. We solve a set of $N_{g}$
linear equations of size $N_{\chi}\times N_{\chi}$ to obtain the
fitting functions $\chi_{\mathbf{R}_{g},\mathbf{R}}$. A naive evaluation
of $X$ will cost $N^{2}N_{g}N_{\xi}$, however, by using the product
structure of $\tilde{\rho}=\tilde{\phi}_{\mu}\tilde{\phi}_{\nu}$
we reduce the cost to $NN_{g}N_{\chi}$ 
\[
X_{\mathbf{R}'_{g},\mathbf{R}}=\left(\tilde{\phi}_{\mu,\mathbf{R}_{g}}\tilde{\phi}_{\mu,\mathbf{R}}\right)^{2}
\]
Thus, by obtaining the interpolation points using QRCP and the fitting
functions by solving the least square problem we have a way of systematically
improving approximation in (\ref{eq:isdf}). We show in Section~\ref{subsec:Robust-fitting}
how to use ISDF to reduce the CPU cost for exchange evaluation.

\subsection{Robust fitting\label{subsec:Robust-fitting}}

Substituting equation (\ref{eq:isdf}) into equation (\ref{eq:2eint}),
we obtain a set of approximate integrals 
\begin{align}
(\mu\nu|\lambda\sigma) & \approx\frac{1}{\Omega}\sum_{\mathbf{R}_{g}}\sum_{\mathbf{R}\mathbf{R}'}\tilde{\phi}_{\mu}(\mathbf{R}_{g})\tilde{\phi}_{\nu}(\mathbf{R}_{g})\left(\chi_{g}(\mathbf{R})\sum_{\mathbf{G}}\exp(-i\mathbf{G}\cdot\mathbf{R})\frac{4\pi}{G^{2}}\exp(i\mathbf{G}\cdot\mathbf{R}')\right)\tilde{\phi}_{\lambda}(\mathbf{R}')\tilde{\phi}_{\sigma}(\mathbf{R}')+O(\epsilon)\nonumber \\
 & \approx\frac{1}{\Omega}\sum_{\mathbf{R}_{g}}\sum_{\mathbf{R}'}\tilde{\phi}_{\mu}(\mathbf{R}_{g})\tilde{\phi}_{\nu}(\mathbf{R}_{g})V(\mathbf{R}_{g},\mathbf{R}')\tilde{\phi}_{\lambda}(\mathbf{R}')\tilde{\phi}_{\sigma}(\mathbf{R}')+\mathcal{O}\left(\epsilon\right)\label{eq:rpsPot}
\end{align}
where we have defined $V(\mathbf{R}_{g},\mathbf{R})$ as the potential
due to the function $\chi_{g}(\mathbf{R})$. However, these integrals
are not symmetric with respect to functions on the bra and ket (i.e.
$(\mu\nu|\lambda\sigma)\neq(\lambda\sigma|\mu\nu)$), and consequently
lead to exchange matrix that is asymmetric. The integrals can be made
symmetric and more importantly the error can be made $\mathcal{O}\left(\epsilon^{2}\right)$
(which is a significant improvement over $O(\epsilon)$ error) by
using the following modification 
\begin{align}
(\mu\nu|\lambda\sigma)\approx & \frac{1}{\Omega}\sum_{\mathbf{R}_{g}}\sum_{\mathbf{R}'}\tilde{\phi}_{\mu}(\mathbf{R}_{g})\tilde{\phi}_{\nu}(\mathbf{R}_{g})V(\mathbf{R}_{g},\mathbf{R}')\tilde{\phi}_{\lambda}(\mathbf{R}')\tilde{\phi}_{\sigma}(\mathbf{R}')\nonumber \\
 & +\frac{1}{\Omega}\sum_{\mathbf{R}}\sum_{\mathbf{R}_{g}}\tilde{\phi}_{\mu}(\mathbf{R})\tilde{\phi}_{\nu}(\mathbf{R})V(\mathbf{R}_{g},\mathbf{R})\tilde{\phi}_{\lambda}(\mathbf{R}_{g})\tilde{\phi}_{\sigma}(\mathbf{R}_{g})\nonumber \\
 & -\frac{1}{\Omega}\sum_{\mathbf{R}'_{g}}\sum_{\mathbf{R}_{g}}\tilde{\phi}_{\mu}(\mathbf{R}_{g})\tilde{\phi}_{\nu}(\mathbf{R}_{g})W(\mathbf{R}_{g},\mathbf{R}'_{g})\tilde{\phi}_{\lambda}(\mathbf{R}'_{g})\tilde{\phi}_{\sigma}(\mathbf{R}'_{g})+\mathcal{O}\left(\epsilon^{2}\right)\label{eq:rps}
\end{align}
where $W(\mathbf{R}_{g},\mathbf{R}'_{g})$ is a symmetric matrix given
by 
\begin{equation}
W(\mathbf{R}_{g},\mathbf{R}'_{g})=\sum_{\mathbf{R}}V(\mathbf{R}_{g},\mathbf{R})\chi_{g}(\mathbf{R})\label{eq:thcpot}
\end{equation}
To see where the qudratic error in \eqref{eq:rps} comes from, it
is useful to recognize that the two electron integral is of the form
$\sum_{ab}v(a)M(a,b)w(b)$. Now we can approximate $v(a)=\sum_{a_{g}}v(a_{g})\xi_{g}(a)+\epsilon\delta(a)$
and write a similar expression for $w(b)$, where $\delta(a)$ is
a normalized error vector and $\epsilon$ is included to signify the
magnitude of the error. Using these approxiamtions one can write 
\begin{align*}
\sum_{ab}v(a)M(a,b)w(b)= & \sum_{ab}\text{\ensuremath{\left(\sum_{a_{g}}v(a_{g})\xi_{g}(a)+\epsilon\delta(a)\right)}}M(a,b)\left(\sum_{b_{g}}w(b_{g})\xi_{g}(b)+\epsilon'\delta'(b)\right)\\
= & \sum_{a'b}v(a_{g})\text{\ensuremath{\left(\sum_{a}\xi_{g}(a)M(a,b)\right)}}w(b)+\sum_{ab'}v(a)\text{\ensuremath{\left(\sum_{b}M(a,b)\xi_{g}(b)\right)}}w(b_{g})\\
 & +\sum_{a'b'}v(a_{g})\text{\ensuremath{\left(\sum_{ab}\xi_{g}(a)M(a,b)\xi_{g}(b)\right)}}w(b)+\epsilon\epsilon'\sum_{ab}\delta(a)M(a,b)\delta'(b)
\end{align*}
The above expression is exact and in the robust pseudospectral (rPS)
technique we include the first three terms and exclude the final term
which is quadratic in the error $\epsilon$. The approximation made
in equation (\ref{eq:rps}) is known as the robust pseudospectral
(rPS) and was introduced recently \citep{Pierce2021}. If one only
includes the first line of the equation then we get the pseudospectral
(PS) method and if one only uses the last line of the equation then
we get the tensor hyper-contraction (THC) method. Recently, it was
emphasized by Valeev et al \citep{Pierce2021} that the quadratic
error can be obtained by using rPS. In fact this approach has been
previously used in other contexts including robust density fitting
\citep{Dunlap2000a,Dunlap2000} and even in quantum Monte Carlo for
the evaluation of the reduced density matrices \citep{Mahajan2022,whitlock1979properties,ceperley1986quantum,rothstein2013survey}.

The key point is that one can evaluate the potential due to the functions
$\chi_{g}(\mathbf{R})$ only at the beginning of the calculation and
then during subsequent evaluations of the exchange matrix one can
avoid having to do FFT. Below we show the order in which the tensor
contraction can be performed to obtain the exchange matrix 
\begin{equation}
K_{\mu\nu}=\frac{1}{\Omega}\sum_{\mathbf{R}_{g}}\tilde{\phi}_{\mu}(\mathbf{R}_{g})\left(\sum_{\mathbf{R}'}V(\mathbf{R}_{g},\mathbf{R}')\left(\sum_{i}\left(\sum_{\sigma}C_{\sigma i}\tilde{\phi}_{\sigma}(\mathbf{R}_{g})\right)\left(\sum_{\lambda}C_{\lambda i}\tilde{\phi}_{\lambda}(\mathbf{R}')\right)\right)\tilde{\phi}_{\nu}(\mathbf{R}')\right),\label{eq:rpsEx}
\end{equation}
where we have only focused on the first term of equation (\ref{eq:rps})
(other terms can be treated in a similar way). More explicitly, the
steps involved are 
\begin{enumerate}
\item First contraction over $\lambda$ and $\sigma$ indices is carried
out to obtain molecular orbitals at all grid points ($\tilde{\phi}_{i}(\mathbf{R}')$)
and interpolation grid points ($\tilde{\phi}_{i}(\mathbf{R}_{g})$)
respectively. The cost of both these calculations is $O(nN_{g})$
because of the locality of the atomic orbitals. 
\item Next the contraction over $i$ is carried out to obtain the matrix
$\rho(\mathbf{R}_{g},\mathbf{R}')=\sum_{i}\tilde{\phi}_{i}(\mathbf{R}')\tilde{\phi}_{i}(\mathbf{R}_{g})$
and the cost of this contraction is $O(nN_{g}N_{\chi})$. 
\item Next we element-wise multiply $V$ and $\rho$ matrices to obtain
$V_{2}(\mathbf{R}_{g},\mathbf{R}')=V(\mathbf{R}_{g},\mathbf{R}')\rho(\mathbf{R}_{g},\mathbf{R}')$
which costs $O(N_{\chi}N_{g})$. 
\item The matrix $V_{2}(\mathbf{R}_{g},\mathbf{R}')$ is contracted with
$\tilde{\phi}_{\mu}(\mathbf{R}_{g})$ to obtain a new matrix $M_{\mu}(\mathbf{R}')=\sum_{\mathbf{R}_{g}}V_{2}(\mathbf{R}_{g},\mathbf{R}')\tilde{\phi}_{\mu}(\mathbf{R}_{g})$
at a cost of $O(NN_{g}N_{\chi})$. 
\item Finally, we contract over $\mathbf{R}'$ to obtain exchange matrix
$K_{\mu\nu}=\sum_{\mathbf{R}'}M_{\mu}(\mathbf{R}')\tilde{\phi}_{\nu}(\mathbf{R}')$
at a cost of $O(N_{g}N^{2})$. 
\end{enumerate}
Given that $N_{g}>N_{\chi}>N>n$ we can see that several steps of
the algorithm are cubic scaling with step 4 being the most expensive.
Notice that no FFT evaluations are involved and in the result section
we will show that this leads to significant speed up of the calculation.

\subsection{occ-RI}

If one is only interested in calculating the total energy or the energy
of the occupied orbitals then it is easy to show that only the rectangular
part of the Fock matrix $F_{i\mu}$ is needed, where $i$ are the
labels of the occupied molecular orbitals and $\mu$ are the atomic
orbitals. This was first introduced by Manzer et al \citep{Manzer2015a}
to reduce the cost of construction of the exchange matrix. Using $F_{i\mu}$
instead of the full $F_{\nu\mu}$matrix during the SCF cycle does
not deteriorate the rate of convergence and the only drawback is that
the virtual orbital energies are not evaluated correctly. However,
this shortcoming can be overcome at the very end of the SCF cycle
by evaluating the full matrix one time. This idea is also at the heart
of the efficiency of the ACE method proposed by Lin Lin \citep{Lin2016},
with the difference being that the Fock matrix is not explicitly constructed
but is written as a sum of outer product of $n$ vectors. This trick
can readily be used with the integrals given in (\ref{eq:rps}) and
we obtain 
\begin{align}
K_{j\nu}= & \sum_{\mathbf{R}'}\left(\sum_{\mathbf{R}_{g}}\tilde{\phi}_{j}(\mathbf{R}_{g})V(\mathbf{R}_{g},\mathbf{R}')\left(\sum_{i}\left(\sum_{\sigma}C_{\sigma i}\tilde{\phi}_{\sigma}(\mathbf{R}_{g})\right)\left(\sum_{\lambda}C_{\lambda i}\tilde{\phi}_{\lambda}(\mathbf{R}')\right)\right)\right)\tilde{\phi}_{\nu}(\mathbf{R}')\nonumber \\
 & +\sum_{\mathbf{R}_{g}}\left(\sum_{\mathbf{R}}\tilde{\phi}_{j}(\mathbf{R})V(\mathbf{R}_{g},\mathbf{R})\left(\sum_{i}\left(\sum_{\sigma}C_{\sigma i}\tilde{\phi}_{\sigma}(\mathbf{R})\right)\left(\sum_{\lambda}C_{\lambda i}\tilde{\phi}_{\lambda}(\mathbf{R}_{g})\right)\right)\right)\tilde{\phi}_{\nu}(\mathbf{R}_{g})\nonumber \\
 & -\sum_{\mathbf{R}_{g}'}\left(\sum_{\mathbf{R}_{g}}\tilde{\phi}_{j}(\mathbf{R}_{g})W(\mathbf{R}_{g},\mathbf{R}_{g}')\left(\sum_{i}\left(\sum_{\sigma}C_{\sigma i}\tilde{\phi}_{\sigma}(\mathbf{R}_{g})\right)\left(\sum_{\lambda}C_{\lambda i}\tilde{\phi}_{\lambda}(\mathbf{R}_{g}')\right)\right)\right)\tilde{\phi}_{\nu}(\mathbf{R}_{g}')\label{eq:exoccri}
\end{align}
The cost of performing contractions in the first line is $O(nN_{\chi}N_{g})$,
of second line is $O(nN_{\chi}N_{g})$ and third line is $O(nN_{\chi}^{2})$
which are all lower than the leading cost of $O(NN_{g}N_{\chi})$
of evaluating the exchange matrix without occ-RI. However, as we will
see in the next section, when performing the calculation in parallel,
the cost of the second step shown above is either $O(nN_{\chi}N_{g}/k)+O(nN_{\chi}N)$
or $O(NN_{g}N_{\chi}/k)$, where $k$ is the number of processors
(assuming we do not want to incur additional communication overhead)
and the choice between the two is not always obvious. In fact if one
goes with the second choice then occ-RI provides no advantage.

\section{Computational details\label{sec:Computational-details}}

Having outlined the basic algorithm for fast evaluation of exchange
we will outline the computational details of the program paying particular
attention to the way in which it is parallelized by making use of
both the message passing interface (MPI) and OpenMP (OMP) together.
The summary of the memory and CPU cost of the various steps of the
algorithm are displayed in Table \ref{tab:The-table-shows} and the
details are presented in the section specified in the fourth column.

\begin{table}[h]
\begin{tabular}{ccccccc}
\hline 
Steps  & \,\,  & Memory  & \,\,  & CPU & \,\,  & Section\tabularnewline
\hline 
$\phi_{\mu}(\mathbf{R})$  &  & $O(N_{g}/k)$  &  & $O(NN_{g}/k)$ &  & \ref{subsec:Voronoi-partitioning-for}\tabularnewline
$J_{\mu\nu}$  &  & $O(N^{2})$  &  & $O(N_{g}/k)$ &  & \ref{subsec:Parallel-Coulomb-matrix}\tabularnewline
$\chi_{g}(\mathbf{R})$  &  & $O(N_{\chi}N_{g}/k)$  &  & $O(N_{\chi}^{3}/k_{t})$ &  & \ref{subsec:ISDF}\tabularnewline
$V(\mathbf{R}_{g},\mathbf{R}')$  &  & $O(N_{\chi}N_{g}/k)$  &  & $O((N_{\chi}N_{g}\ln N_{g})/k)$ &  & \ref{subsec:ISDF}\tabularnewline
$W(\mathbf{R}_{g},\mathbf{R}'_{g})$  &  & $O(N_{\chi}^{2}/k)$  &  & $O(N_{\chi}^{2}N_{g}/k)$ &  & \ref{subsec:ISDF}\tabularnewline
$K_{\mu\nu}$  &  & $O(N^{2})$  &  & $O(nN_{\chi}N_{g}/k)+O(nN_{\chi}N)$ &  & \ref{subsec:Parallel-exchange-matrix}\tabularnewline
\hline 
\end{tabular}

\caption{The table shows the memory and cpu cost for constructing the various
tensors in the algorithm. All the symbols are defined in Table \eqref{tab:Notation-that-will}.
For details, please see the text in the sections pointed out in the
last column.\label{tab:The-table-shows}}
\end{table}

\subsection{Voronoi partitioning for each atom\label{subsec:Voronoi-partitioning-for}}

We begin by partitioning the grid points $\{\mathbf{R}\}$ into disjoint
set of points $\kappa_{I}$, one set for each nucleus $I$, such that
all points in the set are closer to atom $I$ (or its periodic images)
than any other atom (or its periodic images). The algorithm for doing
Voronoi partitioning for periodic unit cells is standard \citep{https://doi.org/10.48550/arxiv.2011.00367}.
The sets are distributed in a round robin fashion between different
processors and roughly an equal number of grid points end up on each
processor.

After the partitioning is performed, each processor is used to evaluate
the value of periodized Gaussian basis on the grid point belonging
to $\kappa_{I}$ associated with it. These values, up to a user-defined
threshold $\epsilon_{1}$ (usually $10^{-8}$), are stored in memory.
Thus we end up getting a different matrix $\tilde{\phi}_{\mu}^{I}(\mathbf{R})$
for each atom $I$, which specifies the values of only those functions
$\mu$ that have a non-negligible value on a subset of grid points
$\mathbf{R}$ in the Voronoi partition $\kappa_{I}$. The memory cost
of storing this matrix is $O(1)$ because only those $\mu$ are included
that have a value above $\epsilon_{1}$ for at least one grid point
in $\kappa_{I}$ and due to the local nature of Gaussians, this only
happens for a number of functions that are asymptotically system-size
independent. The number of grid points $\mathbf{R}$ in $\kappa_{I}$
are also system size independent. Thus the total memory requirement
for storing the Gaussian basis set is linear in system size and it
is equally divided among the different processors, each processor
stores $O(N/k)$ amount of memory. Although in principle one should
be able to calculate the matrix $\tilde{\phi}_{\mu}^{I}(\mathbf{R})$
(the value of all Gaussian basis functions at all grid points) at
an $O(N)$ cost by evaluating the values of the Gaussian basis functions
in real space (each Gaussian has compact support up to a finise threshold).
However, in our algorithm we use an $O(N^{2})$ algorithm, whereby
we evaluate the value of the basis functions in the reciprocal space
(recall that the Fourier transform of a Gaussian is also a Gaussian)
and then we use FFT to evaluate the functions at all grid points.
This procedure is more expensive but one avoids the need for lattice
summations (needed for at least small unit cells) and is more convenient
to implement. This calculatio is split up nearly evenly among the
various processors leading to an asymptotic $O(NN_{g}/k)$ cost per
processor. The cost of this step is small enough compared to the rest
of the steps that even for large systems it does not become a bottleneck.

\subsection{Parallel Coulomb matrix formation\label{subsec:Parallel-Coulomb-matrix}}

With the Gaussians stored in memory the three steps of the construction
of the Coulomb matrix are evaluated in parallel on each processor.
The density matrix $D_{\mu\nu}$ is replicated on each processor and
first the density $\rho(\mathbf{R}')=\sum_{\lambda\sigma}D_{\lambda\sigma}\tilde{\phi}_{\lambda}(\mathbf{R}')\tilde{\phi}_{\sigma}(\mathbf{R}')$
is calculated on each processor separately only for the grid points
associated with it. The current algorithm is naturally linear scaling
because only those basis $\lambda,\sigma$ are contracted that have
a non-negligible contribution to the same Voronoi partition $\kappa_{I}$.
The cumulative density is then evaluated by summing up the contributions
to density coming from all processors by using  the MPI command MPI\_Allreduce,
after which the potential is evaluated by calling FFT two times. The
calls to FFT are only parallelized using OMP locally on each processor.
The cost of these FFT calculations is negligible and constitute a
very small fraction of the overall cost and thus not carrying it out
in parallel does not cause significant overhead. Finally after the
potential is obtained, the third step in which the potential is contracted
with the Gaussians on the local grid to obtain the Coulomb operator
$J_{\mu\nu}=\sum_{\mathbf{R}}\tilde{\phi}_{\mu}(\mathbf{R})\tilde{\phi}_{\nu}(\mathbf{R})V(\mathbf{R}')$
is carried out on each processor and then reduced together. Thus one
calculates those elements of the Coulomb operator $J_{\mu\nu}$ for
which both the indices $\mu,\nu$ have non-negligible values on the
Voronoi partitions associated with a given processor. However, the
entire Coulomb matrix is replicated on each processor. Thus the CPU
cost of the algorithm is $O(N_{g}/k)$ and the memory cost is $O(N^{2})$.

The Voronoi partitioning allows us to screen the overlap between Gaussians
efficiently and results in a linear scaling algorithm that is distributed
over different processors almost ideally.

\subsection{ISDF\label{subsec:ISDF}}

Next we evaluate the interpolation points and the fitting functions
for use in exchange build. The input to the program is a number $c$
that specifies the number of interpolation points that will be used
as a factor of the number of basis functions e.g. $c=2$ means that
twice as many interpolation points and fitting functions as the number
of basis functions will be used. When the number of fitting functions
becomes equal to $N_{g}$ then one gets the exact result. Calculating
all interpolation points for the entire system scales as $O(N^{2}N_{g}^{2})$
which is prohibitive. So as a first step, a randomized algorithm \citep{Halko2011,Liberty2007}
is used to reduce the cost. In this algorithm one first constructs
two random matrices $G^{1}$ and $G^{2}$ with orthogonal columns
of size $N\times p$, where $p=\sqrt{N_{\chi}}+m$, where we round
up $\sqrt{N_{\chi}}$ and $m$ is a small integer usually less than
5. Using these matrices we first construct the randomized density
matrix 
\[
\tilde{\rho}_{mn,\mathbf{R}}=\left(\sum_{\mu}G_{\mu,m}^{1}\phi_{\mu}(\mathbf{R})\right)\left(\sum_{\nu}G_{\nu,m}^{2}\phi_{\nu}(\mathbf{R})\right)
\]
where the size of the cumulative index $mn\sim N_{\chi}$, i.e. the
number of fitting bases that will be used. The QRCP decomposition
is then performed on this much smaller matrix $\tilde{\rho}_{mn,\mathbf{R}}$
instead of the full matrix $\tilde{\rho}_{\mu\nu,\mathbf{R}}$ which
reduces the scaling to $O(N_{\chi}^{2}N_{g})$. The step is still
fairly slow and without additional simplification can dominate the
overall cost of the calculation even for small systems. Previously
Dong et al. \citep{Dong2018} have proposed to use a method based
on centroid voronoi tessellation (CVT) with a weighted K-Mean algorithm
to reduce the cost to $O(N_{\chi}N_{g})$. Here we follow a different
approach. We first calculate a small subset of fitting points for
each Voronoi partitioning $\kappa_{I}$ (described in section \ref{subsec:Voronoi-partitioning-for})
for each atom $I$. If the number of atom centered basis functions
on an atom $I$ is $N_{I}$ then we obtain a set of $\sim cN_{I}+10$
fitting points from each partition by using the same randomized QRCP
as described above. Each of these calculations is extremely fast and
independent of the size of the system, because both the number of
grid points in $\kappa_{I}$ and the number of basis functions $N_{I}$
are small and system size independent. In the second step all these
fitting points are accumulated to form a set $\{\mathbf{R}_{h}\}$
where the number of points is of similar number as $N_{\chi}$ but
much smaller than $N_{g}$. Now we try to find $N_{\chi}$ interpolation
points from all these $\{\mathbf{R}_{h}\}$ grid points again by using
the same randomized algorithm described above, only this time we end
up with a much smaller matrix 
\[
\tilde{\rho}_{mn,\mathbf{R}_{h}}=\left(\sum_{\mu}G_{\mu,m}^{1}\phi_{\mu}(\mathbf{R}_{h})\right)\left(\sum_{\nu}G_{\nu,m}^{2}\phi_{\nu}(\mathbf{R}_{h})\right)
\]
on which one can readily perform QRCP decomposition at a cost of $O(N_{\chi}^{3})$.
The QRCP is parallelized using OMP and thus the CPU cost is $O(N_{\chi}^{3}/k_{t})$,
where $k_{t}$ is the number of threads per node. The construction
of the interpolation points is no longer the dominant cost of the
overall calculation unless one goes to very large system sizes (see
Table \ref{tab:All-calculations-are}). In our current implementation
the QRCP is only parallelized using OMP and not using MPI+OMP. In
a future publication we will use Scalapack\citep{Scalapack} to further
reduce the scaling from $O(N_{\chi}^{3}/k_{t})$ to $O(N_{\chi}^{3}/k)$.

Once the interpolation points have been obtained one can now calculate
the fitting function $\chi_{g}(\mathbf{R})$ by solving (\ref{eq:lsqfit}).
It is worth noting that it contains $N_{g}$ equations which can be
solved in parallel to obtain $\chi_{g}(\mathbf{R})$ with only a subset
of $\mathbf{R}$ that are associated with the processor. Thus the
CPU cost of this step is $O(N_{\chi}^{3}/k_{t})+O(NN_{\chi}N_{g}/k)$.
The first term arises because of the need to perform Cholesky decomposition
of the matrix $X_{\mathbf{R}'_{g},\mathbf{R}{}_{g}}$ and the term
arises because of the need to solve \eqref{eq:lsqfit}. Following
this one needs to obtain $W(\mathbf{R}_{g},\mathbf{R}'_{g})$ using
(\ref{eq:thcpot}) and to do that one has to first obtain the potential
$V(\mathbf{R}_{g},\mathbf{R})$ due to each function $\chi_{g}(\mathbf{R})$
(see (\ref{eq:rpsPot})). If we view $\chi_{g}(\mathbf{R})$ as a
matrix of size $N_{\chi}\times N_{g}$ then after solving the linear
equations we obtain a matrix of size roughly $N_{\chi}\times\left(N_{g}/k\right)$
on each processor where we have $k$ processors, i.e. each processor
contains a subset of columns of the entire matrix. To obtain the potential
we first use MPI\_Alltoall to obtain matrices of size $\left(N_{\chi}/k\right)\times N_{g}$
i.e. each processor now only contains a subset of rows of the entire
matrix. FFT is performed in parallel on these rows to obtain the potential
$V(\mathbf{R}_{g},\mathbf{R})$ where each processor again only retains
a sub-matrix of size $\left(N_{\chi}/k\right)\times N_{g}$ and this
results in a CPU cost of $O(N_{\chi}N_{g}\ln(N_{g})/k)$. After this
a call to MPI\_Alltoall is used to now distribute the matrix $V(\mathbf{R}_{g},\mathbf{R})$
with a column-wise split such that each processor ends up with a sub-matrix
of size $N_{\chi}\times\left(N_{g}/k\right)$. Finally, having access
to $V(\mathbf{R}_{g},\mathbf{R})$ and $\chi_{g}(\mathbf{R})$ one
can evaluate the inner product defined in equation \ref{eq:thcpot}
to obtain $W(\mathbf{R}_{g},\mathbf{R}'_{g})$, which can again be
distributed over all processors leading to a cost of $O(N_{\chi}^{2}N_{g}/k)$.
Finally, in all these calculations the matrices $\chi_{g}(\mathbf{R}),V(\mathbf{R}_{g},\mathbf{R}'),W(\mathbf{R}_{g},\mathbf{R}'_{g})$
are all distributed evenly among the processors and thus require a
memory of $O(N_{\chi}N_{g}/k),O(N_{\chi}N_{g}/k),O(N_{\chi}^{2}/k)$
respectively.

\subsection{Parallel exchange matrix formation\label{subsec:Parallel-exchange-matrix}}

As mentioned in the text below (\ref{eq:exoccri}), when one uses
the occ-RI one has two choices on how to perform contraction in the
second line of the equation. One can first calculate $M_{j}(\mathbf{R}_{g})=\sum_{\mathbf{R}}\tilde{\phi}_{j}(\mathbf{R})V(\mathbf{R}_{g},\mathbf{R})$
followed by $K_{j\nu}=\sum_{\mathbf{R}_{g}}M_{j}(\mathbf{R}_{g})\tilde{\phi}_{\nu}(\mathbf{R}_{g})$
or one can reverse the order of these two contractions. In the first
case, only the first step can be parallelized and the second step
has to be performed on each processor and this leads to the computational
cost $O(nN_{\chi}N_{g}/k)+O(nN_{\chi}N)$, while when the reversed
order is used then the computational cost becomes $O(NN_{g}N_{\chi}/k)$.
In practice the first algorithm tends to be faster unless the number
of processors is very large, because the ratio $N_{g}/n$ can be on
the order of a $10^{3}$ or more.

\subsection{Other considerations}

The nuclear matrix takes the form 
\begin{align*}
N_{\mu\nu} & =\int\frac{-Z_{I}}{|\mathbf{R}_{I}-\mathbf{r}|}\rho_{\mu\nu}(\mathbf{r})d\mathbf{r}\\
 & =\frac{-Z_{I}}{\Omega}\sum_{\mathbf{G}}^{N_{g}}\frac{4\pi}{G^{2}}\exp(i\mathbf{G}\cdot\mathbf{R}_{I})\rho_{\mu\nu}(\mathbf{G})
\end{align*}
where $\rho_{\mu\nu}(\mathbf{G})$ are the Fourier components of the
density $\rho(\mathbf{r})=\phi_{\mu}(\mathbf{r})\phi_{\nu}(\mathbf{r})$
and here we have represented the nucleus as a delta distribution of
charge $Z_{I}$ at position $\mathbf{R}_{I}$. The summation is truncated
at $N_{g}$ and the largest error is incurred when both functions
$\mu,\nu$ are sharp. The expression for the two electron integral
for the same sharp functions is given as 
\[
(\mu\nu|\mu\nu)=\frac{1}{\Omega}\sum_{\mathbf{G}}^{N_{g}}\frac{4\pi}{G^{2}}\left|\rho_{\mu\nu}(\mathbf{G})\right|^{2}
\]
where again the error is due to the use of a finite $N_{g}$. However,
if the nuclear charge is truly assumed to be distributed as a delta
function then it is clear that the error incurred in the two-electron
integrals is quadratic of that of the one incurred while calculating
nuclear integrals. This trend is somewhat tempered due to the fact
that pseudo-potentials used in this work are not quite delta functions,
but nonetheless the nuclear integrals still show the largest error.
To reduce the error, we use an effectively larger $N_{g}$ while calculating
the nuclear integral which has to be done only once and use a sparser
grid for the rest of the calculation (including for the Coulomb and
Exchange matrix formation).

In periodic systems, one often encounters the problem of linear dependencies
of the Gaussian basis sets. This can be understood by looking at the
equation \ref{eq:gaussG}, if the value of $\mu$ is smaller than
$0.3/L^{2}$, where $L$ is the length of the super cell, then one
can confirm that $\hat{\phi}_{\mu}(\mathbf{G})$ is effectively a
constant. For small unit cells, several Gaussian functions in a standard
basis set can become constants and thus are linearly dependent. However
when one tries to reach the thermodynamic limit by increasing $L$
(the dimension of the super cell) then the Gaussians are no longer
going to be linearly dependent and the number of linearly dependent
functions do not increase linearly with the number of super cells
in the system (of course the basis functions corresponding to the
gamma point of the primitive unit cell are still present in the super
cell which are constants). Thus in the large super cell limit we do
not expect the linear dependency problem to be significantly worse
than in a large relatively uniform molecule or a cluster. Nevertheless,
it is true that by using Gaussian basis functions it is difficult
to obtain results in the basis set limit because it is difficult to
systematically improve the basis set without running into linear dependencies
(which also happens for molecular systems). One way to overcome this
is to use a mixed plane wave Gaussian basis set. In our algorithm
it is relatively straightforward to do so such that sharp features
are described by Gaussian basis function and diffuse features are
described by plane wave basis functions. We will pursue this line
of work in the future.

It is well known that the exchange energy has an integrable singularity
which disappears in the infinite systems size limit. However, when
one is using finite sized super cells then the singularity persists
(note the presence of $1/\mathbf{G}^{2}$ in the denominator of equation
\ref{eq:exchange}) and one needs ways of regularizing it. Several
approaches have been proposed \citep{Sundararaman2013,Spencer08}
to do so including using truncated coulomb \citep{Guidon2009} or
the minimum image convention \citep{Tymczak2005,Irmler2018}. Here
we have used a very simple approach whereby we remove the $\mathbf{G}=0$
term in the exchange matrix evaluation and include a correction term
$nM/2$ where $M$ is the Madelung constant of the super-cell. It
is worth pointing out that one can readily use the truncated Coulomb
kernel in exchange evaluation using our algorithm.

\section{Results\label{sec:Results}}

In this section we present calculations using two benchmark systems
Li-H solid (Li$_{4}$H$_{4}$)$_{n}$ and diamond with unit cell (C$_{8}$)$_{n}$.
For both these systems we make use of the Goedecker-Teter-Hutter (GTH)
pseudo-potentials \citep{Hartwigsen98,Goedecker96} and GTH-DZVP and
GTH-TZVP basis sets \citep{Vandevondele2005}. As pointed in the previous
section the nuclear integrals are evaluated using a large $N_{g}$
cutoff (in fact part of the calculation is performed in the real space).
For the calculation of the Coulomb and exchange matrix $N_{g}=35^{3}$
is used for the conventional unit cell. Numerical experiments showed
that this grid is sufficient to deliver an error of less than 0.5
mHartree accuracy in Li-H system and 2 $\mu$Hartree in diamond. In
this section, we first look at the relative accuracy of a single step
randomized ISDF calculation versus Voronoi based one that we have
proposed here. Second, we compare the accuracy of the rPS and THC
approximations for obtaining the exchange energy, the accuracy of
the rPS as we change the basis set and the size of the system. Finally,
we present data on the cost of the calculations for Li-H of various
system sizes, with the largest one being Li$_{256}$H$_{256}$ that
contains 4864 basis functions.

\subsection{Accuracy of approximate ISDF}

\begin{table}[h]
\begin{tabular}{ccccccc}
\hline 
 &  & \multicolumn{2}{c}{Li$_{4}$H$_{4}$} &  & \multicolumn{2}{c}{C$_{8}$(Diamond)}\tabularnewline
\hline 
$c=N_{\chi}/N$  & \,\,  & ISDF  & v-ISDF  & \,\,  & ISDF  & v-ISDF\tabularnewline
\hline 
3  &  & $0.86$  & $0.67$  &  & $44.15$  & $39.15$\tabularnewline
4  &  & $0.18$  & $0.23$  &  & $6.02$  & $6.39$\tabularnewline
5  &  & $0.03$  & $0.08$  &  & $1.15$  & $1.41$\tabularnewline
6  &  & $0.01$  & $0.03$  &  & $0.17$  & $0.29$\tabularnewline
\hline 
\end{tabular}

\caption{Error in $mE_{h}$ for Li$_{4}$H$_{4}$ and C$_{8}$(Diamond) as
$N_{\chi}$ is increased with respect to results obtained with exact
exchange. It is noteworthy that the errors of randomized ISDF (labeled
ISDF) and voronoi-ISDF (the algorithm proposed in this work) are quite
similar. The differences between the two are slightly larger than
the random errors one expects to see due to the use of randomized
algorithm.\label{tab:Error-in-}}
\end{table}

In our algorithm we do not perform a single large ISDF calculation,
instead we first perform a single small ISDF calculation for each
atom and then do a single ISDF calculation to select the most important
pivots points out of the collection of all the pivots points obtained
from the individual calculations (let us call it v-ISDF for Voronoi-ISDF).
Table \ref{tab:Error-in-} shows the error incurred in Li$_{4}$H$_{4}$
and C$_{8}$ systems when the usual randomized ISDF is used as opposed
to v-ISDF. We note that the errors are quite similar pointing to the
fact that the collection of pivot points that we obtain from individual
ISDF calculations contain the important pivot points. One can thus
devise other algorithms to select the most important points out of
this small subset, including for example the K-Means algorithm presented
previously\citep{Dong2018}. We will work on these aspects in a future
publication, particularly because the single ISDF calculation needed
can become expensive for large system sizes as shown in Table \ref{tab:All-calculations-are}.

\subsection{Accuracy of rPS vs THC}

\begin{figure}[h]
\includegraphics[width=0.3\textwidth]{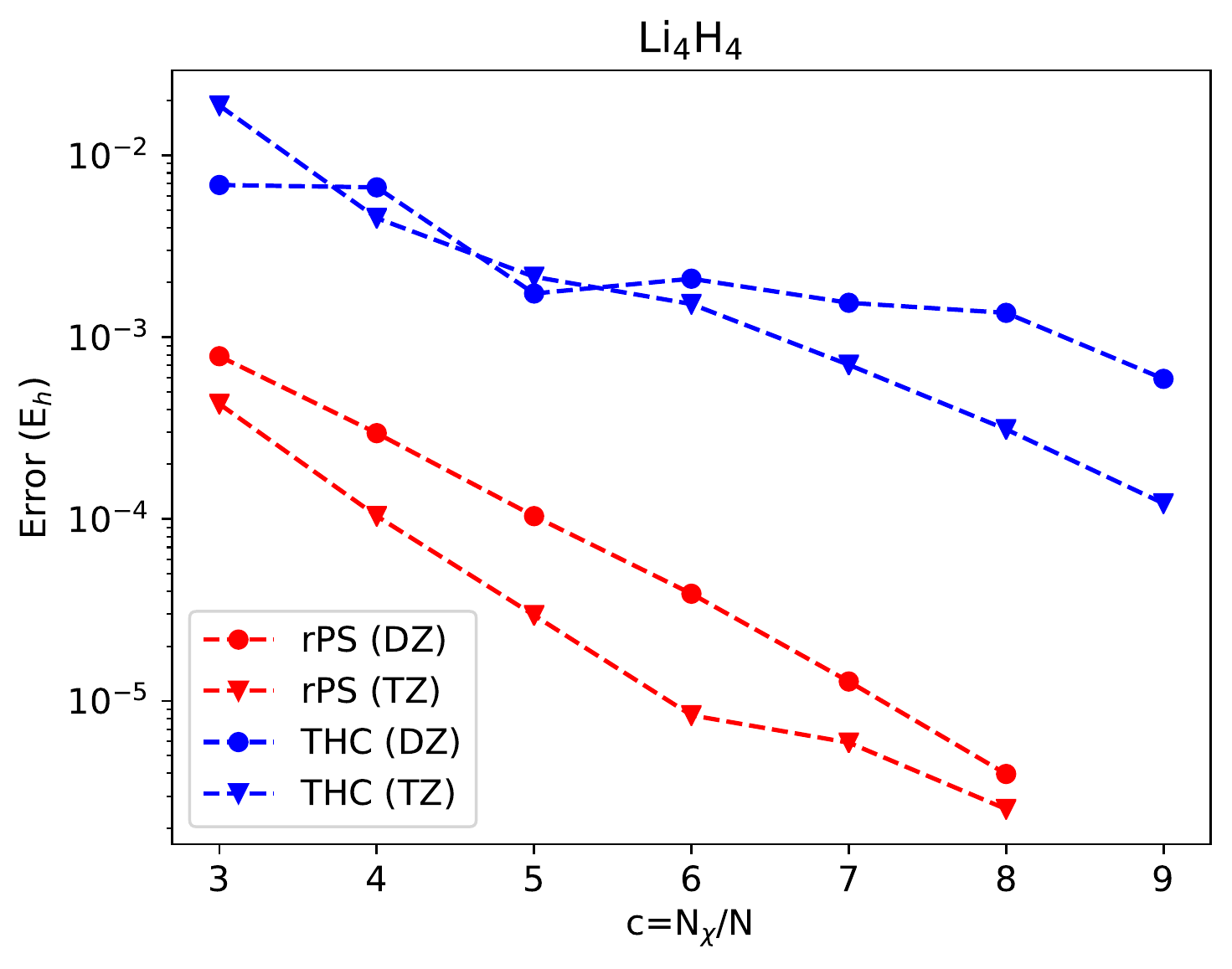}\hfill{}\includegraphics[width=0.3\textwidth]{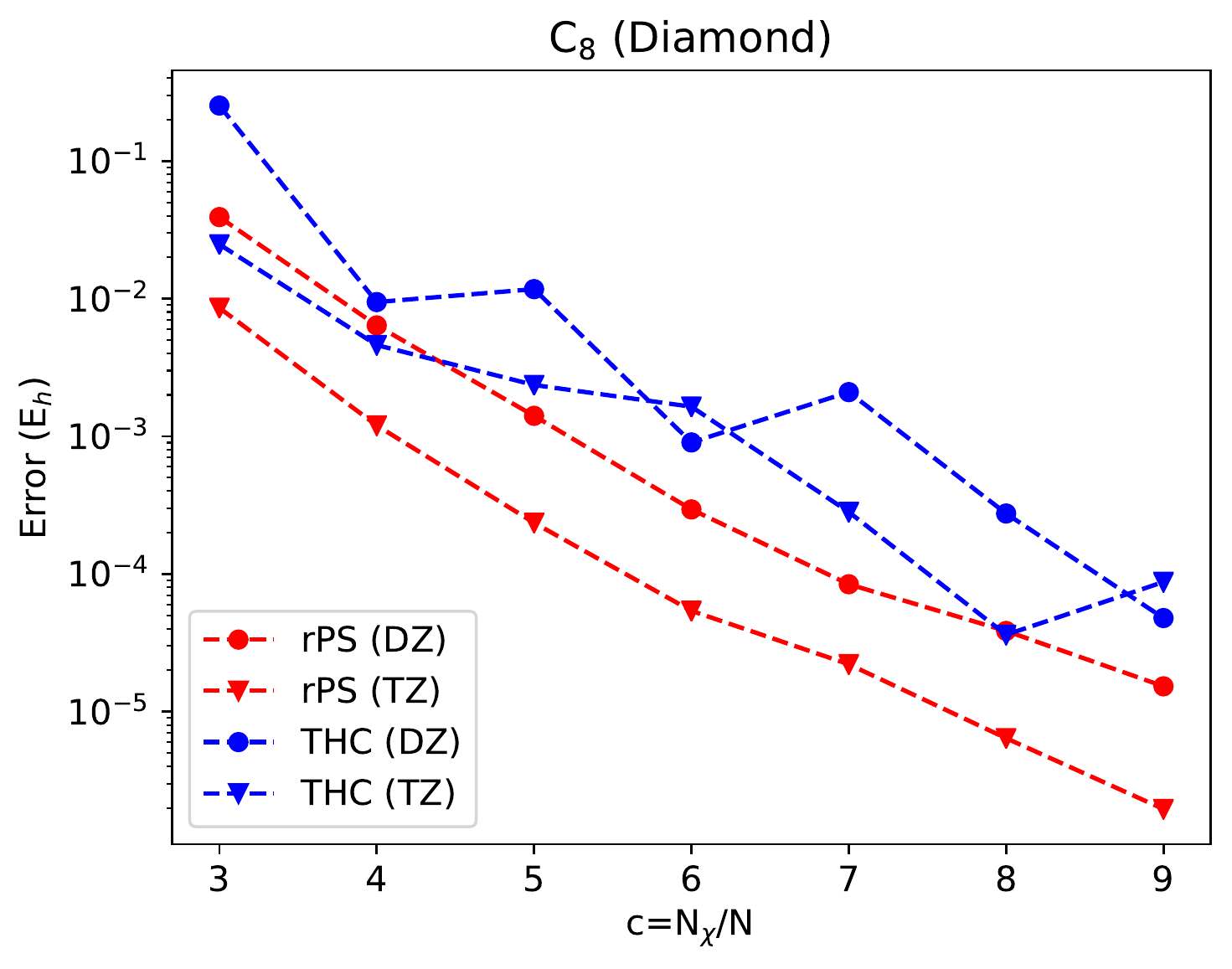}\hfill{}\includegraphics[width=0.3\textwidth]{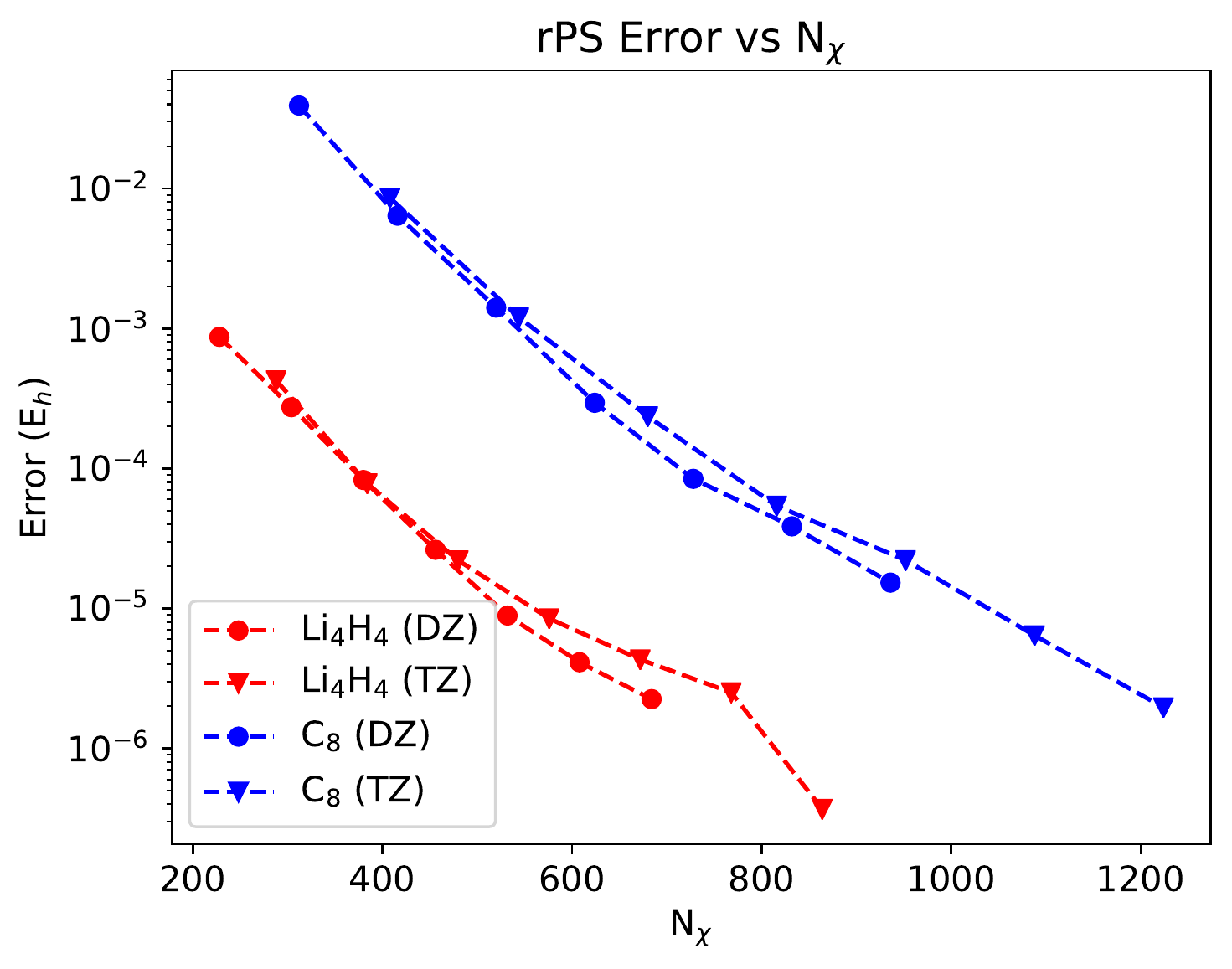}

\caption{The left and the middle figure show the error incurred due to rPS
and THC approximations in Li$_{4}$H$_{4}$ and C$_{8}$(Diamond)
systems respectively. In both systems one sees that the accuracy of
rPS is significantly higher than that of THC and it can be explained
by the quadratic error we expect to get in the former as opposed to
linear error of the latter. A noteworthy observation is that for the
same value of $c$ the error is smaller with the larger basis set
(TZ) rather than the smaller basis set (DZ). In fact, when one plots
the error versus $N_{\chi}$ rather than $c=N_{\chi}/N$ (as is done
is the right most figure) one sees that the curves for the two basis
set nearly overlap, which shows that the error one makes is largely
independent of the size of the basis set for a fixed number of fitting
functions. Another observation is that the error is highly system
dependent (see text and Figure \ref{fig:rdecay}) \label{fig:The-left-and}}
\end{figure}

Figure~\ref{fig:The-left-and} shows the error as a function of $c$
when rPS and THC is used for the two systems. It is clear that the
error incurred by rPS is significantly lower than THC consistent with
the fact that the error in the former is expected to be square of
that of the latter. Two trends stand out in the left two sub-figure.
First, it appears that for the same value of $c$ the error in the
larger TZ basis set is smaller than that in the smaller DZ basis set
and this trend is observed for both systems. Interestingly, when one
plots the error versus $N_{\chi}$rather than $c$ then the errors
in the two curves nearly overlap and again this trend is observed
for both systems (see right most sub-figure). Given that the cost
of formation of the exchange matrix is $O(nN_{\chi}N_{g}/k)+O(nN_{\chi}N)$,
we expect the cost to increase linearly with the increase in the size
of the basis set $N$ (assuming $N_{\chi}$ and $N_{g}$ remain constant).
The second trend one can observe is that the error decreases extremely
rapidly for the Li$_{4}$H$_{4}$ as compared to the C$_{8}$(Diamond).
To understand this fact, in Figure we have plotted the diagonal elements
of the $R$ matrix that is obtained by performing the QR decomposition
of the $\tilde{\rho}_{mn,\mathbf{R}_{h}}$ matrix. These diagonal
elements reflect the importance of the various pivot points and explains
why one has to include a much larger number of pivot points for C$_{8}$(Diamond)
to obtain comparable accuracy. The results show that some amount of
experimentation is needed to decide on the appropriate number of $c$
needed for a desired accuracy. However, the results in Figure~\ref{fig:The-left-and}
show that for both systems the results converge roughly exponentially
fast with the increase in $c$.

\begin{figure}[h]
\includegraphics[width=0.3\textwidth]{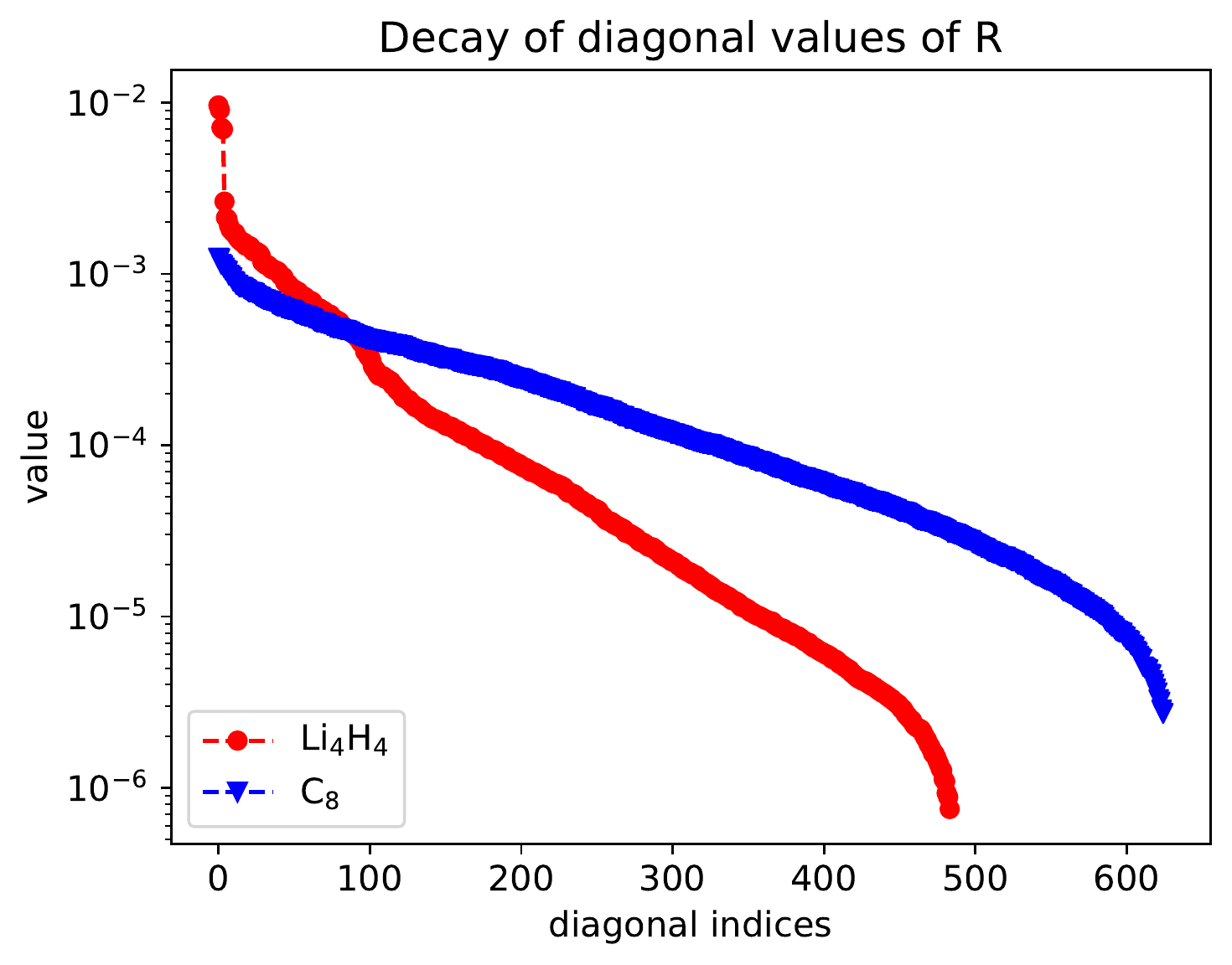}

\caption{The figure shows diagonal elements of the R matrix obtained by performing
the QR decomposition of the $\tilde{\rho}_{mn,\mathbf{R}_{h}}$ matrix.
The diagonal elements decay much faster for Li$_{4}$H$_{4}$ as compared
to the C$_{8}$(Diamond).\label{fig:rdecay}}
\end{figure}

\subsection{Accuracy with system size}

\begin{figure}[h]
\includegraphics[width=0.3\textwidth]{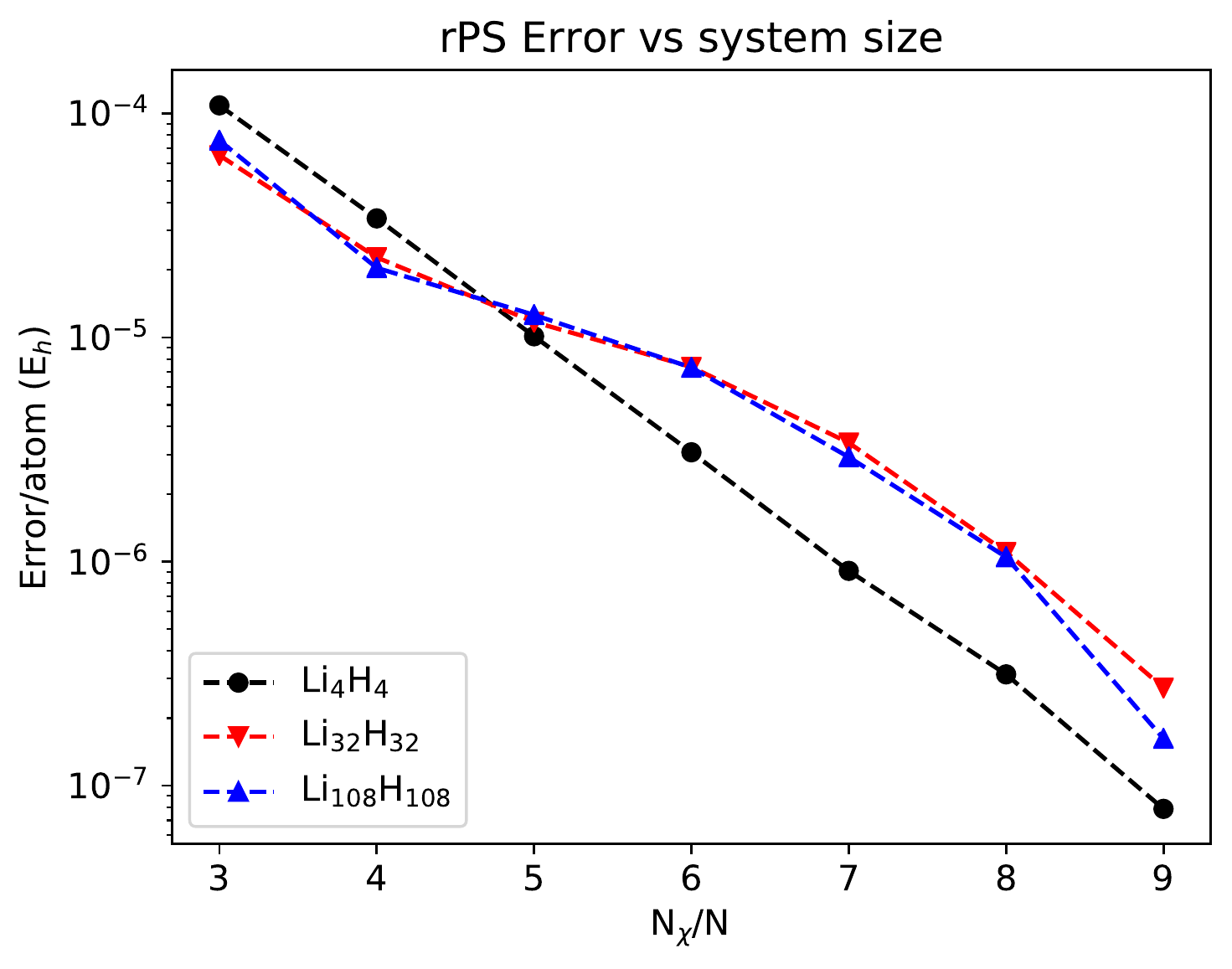}

\caption{The intensive error (Error/atom) is approximately independent of the
size of the system for the same value of $c$. \label{fig:The-intensive-error}}
\end{figure}

In Figure~\ref{fig:The-intensive-error} we have plotted the energy/atom
as a function of $c$ for three systems Li$_{4}$H$_{4}$, Li$_{32}$H$_{32}$
and Li$_{108}$H$_{108}$. The errors/atom are quite similar for the
three systems and in all cases decrease exponentially with $c$. The
figure indicates that one can experiment on a smaller system to estimate
the value of $c$ needed for a desired accuracy and then the calculation
can be scaled to larger system size.

\subsection{Cost of the calculations}

\begin{figure}
\includegraphics[width=0.45\textwidth]{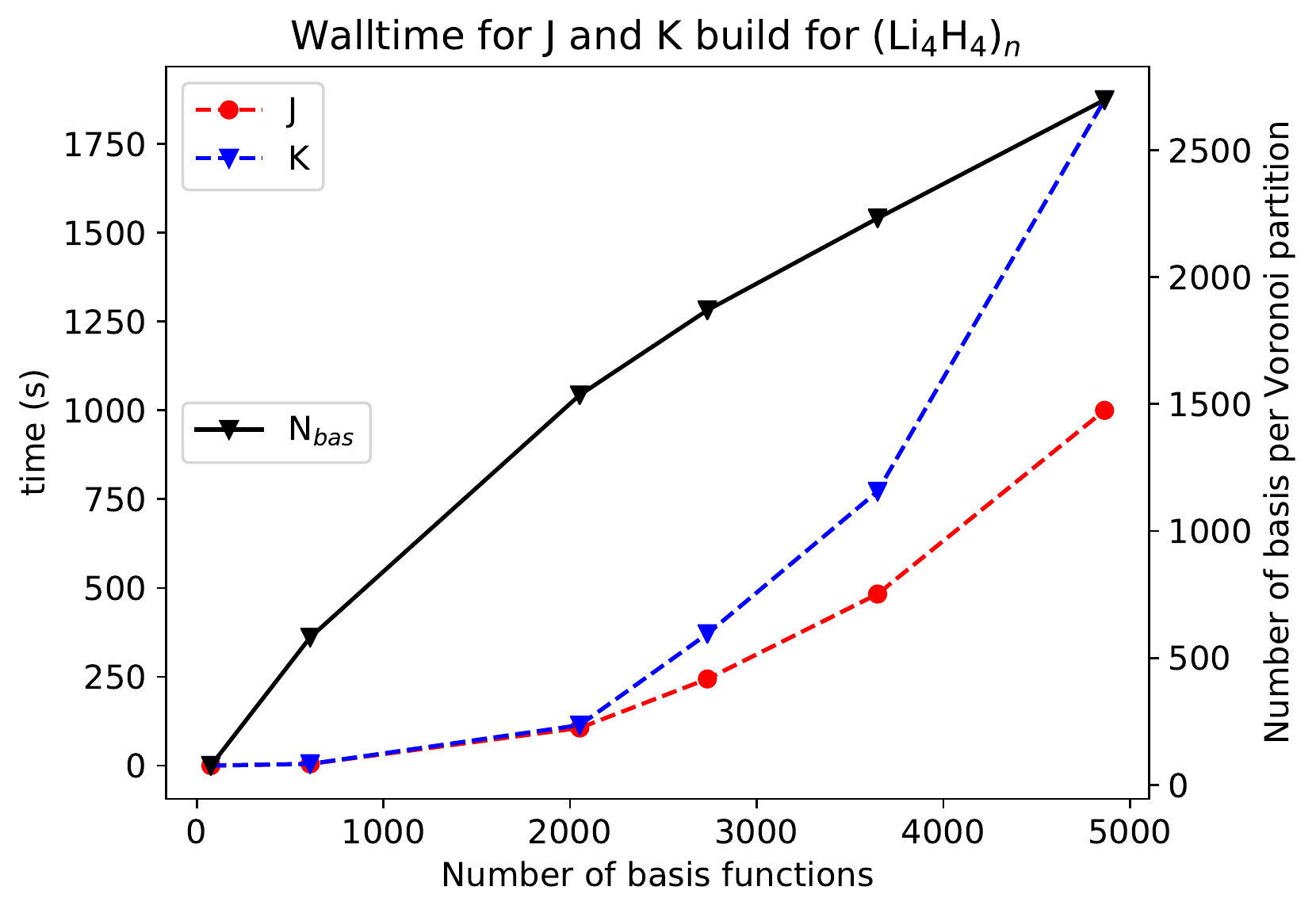}\hfill{}\includegraphics[width=0.45\textwidth]{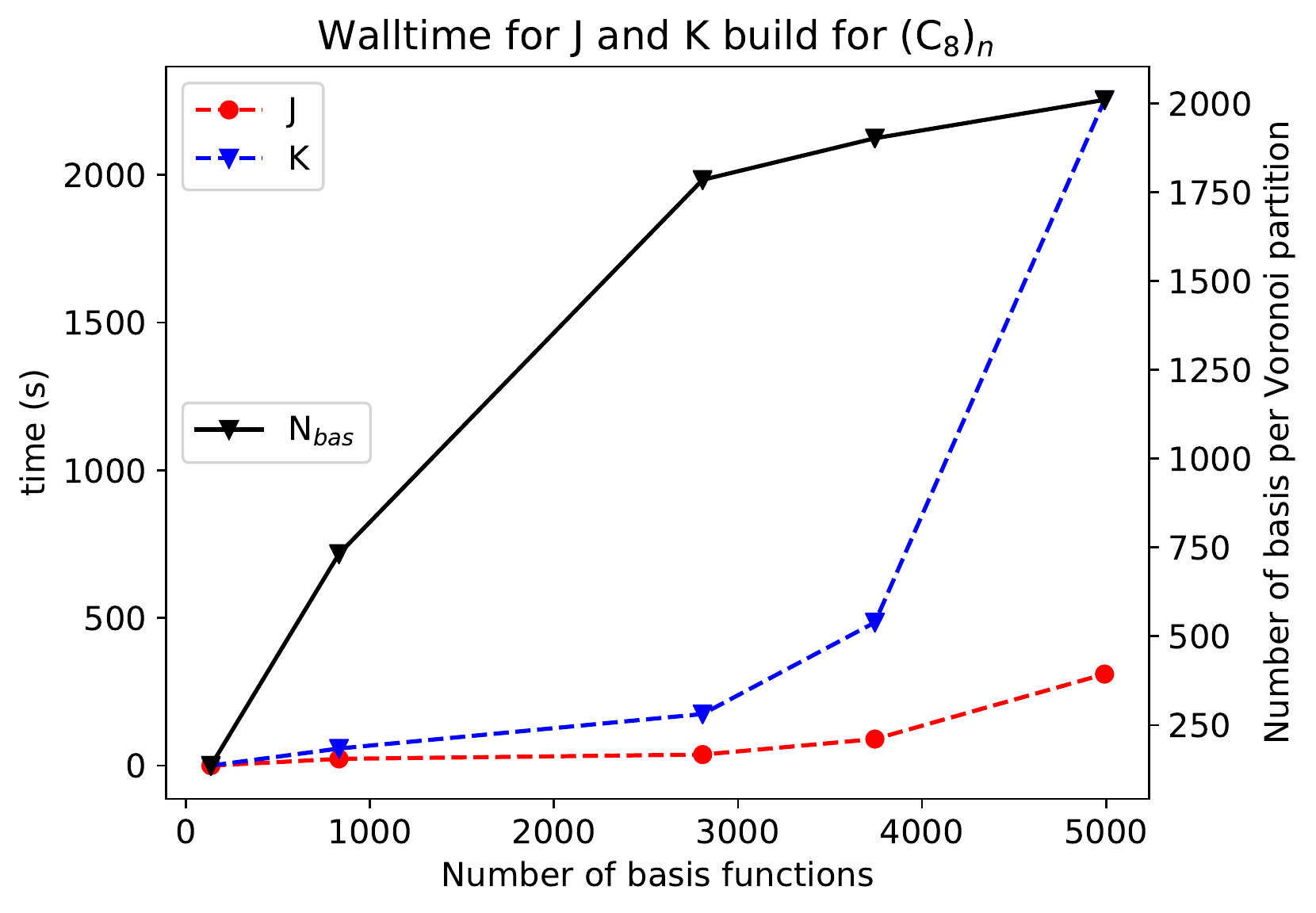}

\caption{The red and blue curve shows the wall time needed for a single build
of the Coulomb and exchange matrix on nodes containing two Intel(R)
Xeon(R) CPU E5-2680 v3 @ 2.50GHz processors. The systems used to obtain
this figure are (Li$_{4}$H$_{4}$)$_{n}$ (lithium hydride) and (C$_{8}$)$_{n}$
(diamond), where $n=1,4,8,27,36,48,64$ for the lithium hydride system
and $n=1,4,8,36,48$ for diamond. For lithium hydride a $c=4$ was
used and for diamond $c=6$ was used. These values of $c$are chosen
to ensure that sub $mE_{h}$ errors are obtained for the $n=1$ calculation.
The timings were normalized for that of 1 node, for example the largest
calculation for lithium hydride contained $64$ unit cells and on
the figures shows a timing of 1000 seconds for Coulomb build, however
the wall time was $1000/16=62.5$ seconds on 16 nodes. The black line
shows the number of basis functions with a value $>10^{-8}$ on the
Voronoi partition. Note that this number is not yet a constant which
leads to the super linear scaling of the Coulomb matrix construction.
The relative cost of exchange is higher in diamond because a higher
value of $c=6$ as opposed to $c=4$ for lithium hydride is used.
Note that the cost of exchange increases linearly with $c$ while
the cost of the Coulomb matrix construction is independent of it.
\label{fig:The-figure-shows}}
\end{figure}

Figure~\ref{fig:The-figure-shows} shows the cost of a single Coulomb
and Exchange build for lithium hydride and diamond for various system
sizes. The scaling of the Coulomb matrix appears to be non-linear
mainly because the system contains highly diffuse functions that do
not decay rapidly enough. For example, the black line in both the
graphs shows that the number of basis functions that have a non-negligible
value on the grid points of the Voronoi partition increases with the
size of the system. Asymptotically this number becomes a constant
but because of the presence of extremely diffuse functions this does
not happen even for the largest system that contains more than a 1000
electrons. In our current implementation the grid spacing is determined
by the sharpest gaussian while the number of basis functions with
non-negligible value is determined by the most diffuse function. This
naturally leads to a suboptimal performance in the Coulomb build.
In the software packages CP2K\citep{Kuhne2020} and PySCF\citep{pyscf2018,pyscf2020}
a technique known and multigrid (different than the multigrid approach
used in the solution of partial differential equation) is used, whereby
several grids are utilized from sparse to dense with sparse grid used
to represent diffuse basis functions and the dense for sharp basis
function. This allows one to reach the linear scaling regime rapidly.
Another option in this context is to use (and further develop) gaussian
basis set recently introduced by Ye and Berkelbach\citep{Ye22} that
have fewer diffuse functions which leads to fewer linear dependency
problems and as a side benefit would allow one to reach the linear
scaling regime more quickly without having to implement multigrid
approaches. Finally, one can also replace diffuse Gaussians with plane
waves which will not only allow us to more systematically increase
the accuracy of the calculation but will also help reduce the overall
cost by reaching the linear scaling regime sooner. We plan to pursue
these lines of work in a future publication.

\begin{table}[h]
\begin{tabular}{rcrcccccccccc}
\hline 
System  & \,\,  & $N$  & \,\,  & K & \,\,  & QRCP\footnote{This is parallelized using only OMP and not MPI.}  & \,\,  & $\chi_{g}(\mathbf{R})$  & \,\,  & $W(\mathbf{R}_{g},\mathbf{R}'_{g})$/(MPI)  & \,\,  & Diagonalize$^{\mathrm{a}}$\tabularnewline
\hline 
Li$_{4}$H$_{4}$  &  & 76  &  & 1.2  &  & 1.0  &  & 0.4  &  & 1.3/(0.0)  &  & 0.0\tabularnewline
Li$_{32}$H$_{32}$  &  & 608  &  & 1.0  &  & 0.3  &  & 0.3  &  & 0.7/(0.0)  &  & 0.0\tabularnewline
Li$_{108}$H$_{108}$  &  & 2052  &  & 1.1  &  & 0.6  &  & 0.4  &  & 0.6/(0.3)  &  & 0.0\tabularnewline
Li$_{144}$H$_{144}$  &  & 2736  &  & 1.4  &  & 0.7  &  & 0.5  &  & 0.7/(0.3)  &  & 0.0\tabularnewline
Li$_{192}$H$_{192}$  &  & 3648  &  & 1.6  &  & 1.8  &  & 0.5  &  & 1.9/(1.4)  &  & 0.3\tabularnewline
Li$_{256}$H$_{256}$  &  & 4864  &  & 1.9  &  & 3.9  &  & 0.6  &  & 2.9/(1.9)  &  & 0.6\tabularnewline
\hline 
\end{tabular}

\caption{In this table we compare the cost of various steps of the algorithm
relative to the cost of the Coulomb matrix construction as the size
of the Lithium hydride system is increased. All calculations are $\Gamma$
point calculations and do not make use of $k$-point symmetry. For
Li$_{4}$H$_{4}$ the various parameters are $n=16,N=76,N_{\chi}=304,N_{g}=42875$
and for other systems each of these parameters increase in proportion
to the size of the system. Column ``K'' referes to the cost of exchange
build, column ``QRCP'' refers the cost of performing of obtaining
the interpolation points, column ``$\chi_{g}(\mathbf{R})$'' refers
to the cost of obtaining the interpolation functions by solving the
least square problem and column $W(\mathbf{R}_{g},\mathbf{R}'_{g})$
refers to the cost of building the matrix $W(\mathbf{R}_{g},\mathbf{R}'_{g})$
in \eqref{eq:rps}. All Hartree-Fock calculations converged with 8
iterations and the CPU cost is relative to the cost for constructing
the Coulomb matrix 8 times. In column 6, the number in brackets represents
the time spent in just the MPI calls. \label{tab:All-calculations-are}}
\end{table}

Table~\ref{tab:All-calculations-are} shows the timings for the various
steps of the calculation for a series of Li-H solids of increasing
supercell size. All calculations are performed using computational
nodes containing two Intel(R) Xeon(R) CPU E5-2680 v3 @ 2.50GHz processors
with a total number of threads equalling 24. In all calculations a
value of $c=4$ was used. From these calculations it is clear that
the cost of performing QRCP to obtain the interpolation points increases
quite rapidly with the systems size. This is particularly true because
the QRCP is not parallelized over the number of processors and only
OMP is used. We are optimistic that by using Scalapack one can reduce
the cost of this step. Besides this step the cost of evaluating $W(\mathbf{R}_{g},\mathbf{R}'_{g})$
also increases rather steeply. A significant part of the cost is incurred
due to the calls to the MPI\_All-to-all function which is given in
the bracket of column 6. The relative cost of evaluation of the exchange
matrix never increases beyond 2 compared to the Coulomb matrix even
for large systems. However, for even larger systems this relative
cost will keep increasing because the scaling of the two steps is
not the same. Number of nodes needed increases quadratically with
the size of the system because the memory cost of storing $V(\mathbf{R}_{g},\mathbf{R})$
is $N_{\chi}N_{g}$, for example 16 nodes were needed to perform the
largest calculation in Table~\ref{tab:All-calculations-are}.

\section{Conclusions}

In this work we have presented an algorithm for reducing the cost
of evaluation of the exchange matrix such that it is only slighly
more expensive than the evaluation of the Coulomb matrix. We have
done so by reducing the prefactor without changing the scaling. Currently
the most expensive step of the algorithm is obtaining the interpolation
points using QRCP which is cubic scaling with a fairly large over
head. The QRCP cost can be decreased in two ways. First, one can parallelize
the current algorithm using Scalapack to effectively make use of both
the MPI and OMP, while currently we only make use of OMP. Second,
we can reduce the computational cost of the algorithm as follows.
In the first step of the current algorithm we obtain a set of interpolation
points from each atom and in the second step we perform a large QRCP
calculation to identify the most important points in this set. The
second step is by far the dominant cost and this can be replaced by
selecting points by using a criterion different than the QRCP method.
For example, the centroid Voronoi tesselation (CVT) with K-mean algorithm
\citep{Dong2018} can be used to replace the last step. We expect
this to be an effective approach because CVT is only used for selecting
a subset of points which are already close to optimal.

In addition, the algorithm can be extended in several ways. First,
we can use a mixture of plane wave and Gaussian basis function by
removing diffuse Gaussians and replacing them with a set of plane
wave basis. By increasing the number of plane wave basis we expect
to be able to reach basis set limit when calculating energy differences
(e.g. atomization energies). It remains to be seen how effective rPS
algorithm remains at selecting pivot points to represent the product
density of this mixed basis set. Also the number of basis functions
will increase rapidly with the threshold and one will most likely
have to resort to direct diagonalization. Second, our algorithm has
similarities with the discontinuous Galerkin method \citep{McClean2020,https://doi.org/10.48550/arxiv.2011.00367}
and it is possible to obtain a set of Galerkin basis that are localized
to a given Voronoi partition. This will have a significant advantage
that the fitting functions will also be perfectly localized to an
atomic domain and thus the cubic scaling of the QRCP step and the
evaluation of the $W(\mathbf{R}_{g},\mathbf{R}'_{g})$ will be reduced
to linear and quadratic scaling respectively. Third, we can implement
$k$-point symmetry which is expected to reduce the cost of the exchange
evaluation to $O(n_{\mathbf{k}}\ln(n_{\mathbf{k}}))$ where $n_{\mathbf{k}}$
is the number of $k$-points\citep{Jinlong2022}. Fourth, if one wants
to avoid the use of pseudo-potential then sharp gaussians are needed.
To include such Gaussians one can split the solution of the Poisson's
equation between real space and reciprocal space \citep{sharmaBeylkina},
with real space calculations requiring the evaluation of mixed-Gaussians-plane
wave integrals which we have recently developed\citep{Beylkin2021}.
Alternatively, one has to use an irregular grid, where wavelet basis
with multi resolution analysis is an attractive approach. Several
wavelet basis such as interpolating wavelets, Coiflets and Gausslets
allow one to use the diagonal approximation that is needed for our
algorithm to work. Finally, it is possible to use the rPS integrals
in correlated calculations for periodic and molecular systems.

\section{Acknowledgements}

AFW and SS were partly supported by the DOE grant DE-SC0022385. SS
was also partly supported by NSF Career award CHE-2145209. We would
also like to thank Toru Shiozaki and Daniel Moberg for helpful discussions.

%\bibliographystyle{unsrt}
%\bibliography{refs}

\end{document}